\documentclass[draftclsnofoot,onecolumn]{IEEEtran}

\usepackage{amssymb}
\usepackage{amsmath}
\usepackage{graphicx}
\usepackage{color}	

\newcommand{\dv}{d_\mathrm{v}}
\newcommand{\dc}{d_\mathrm{c}}
\newcommand{\GF}{\mathrm{GF}}

\begin{document}

%

\title{Approaching the Capacity of Large-Scale MIMO Systems via Non-Binary LDPC Codes}

%
%
%

\author{Puripong~Suthisopapan,~\IEEEmembership{Student Member,~IEEE,}
        Kenta~Kasai,~\IEEEmembership{Member,~IEEE,} \\
        Anupap Meesomboon, 
        and~Virasit Imtawil
        }


\maketitle


\begin{abstract}
In this paper, 
the application of non-binary low-density parity-check (NBLDPC) codes to 
MIMO systems which employ hundreds of antennas at both the transmitter and the receiver has been proposed.
Together with the well-known low-complexity MMSE detection,
the moderate length NBLDPC codes can operate closer to the MIMO capacity, 
e.g., capacity-gap about 3.5 dB  (the best known gap is more than 7 dB).
To further reduce the complexity of MMSE detection,
a novel soft output detection that can provide 
an excellent coded performance in low SNR region
with 99$\%$ complexity reduction is also proposed.
The asymptotic performance is analysed by using the Monte Carlo density evolution.
It is found that the NBLDPC codes can operate within 1.6 dB from the MIMO capacity.
Furthermore, the merit of using the NBLDPC codes in large MIMO systems with
the presence of imperfect channel estimation and spatial fading correlation
which are both the realistic scenarios for large MIMO systems is also pointed out.
\end{abstract}

\begin{IEEEkeywords}
Large MIMO systems, Hundreds of antennas, Non-binary LDPC codes, MMSE detection, MF detection, Soft-output detection.
\end{IEEEkeywords}
\let\thefootnote\relax\footnote{P. Suthisopapan, A. Meesomboon*, and V. Imtawil are with the Department of Electrical Engineering, 
Khon Kaen University, Thailand (e-mail:
mr.puripong@gmail.com; anupap@kku.ac.th; virasit@kku.ac.th) (*corresponding author).}
\let\thefootnote\relax\footnote{K. Kasai is with the Department of Communications and Integrated Systems, 
Graduate School of Science and Engineering, Tokyo Institute of Technology, Tokyo 152-8550, Japan 
(e-mail:kenta@comm.ss.titech.ac.jp).}
\let\thefootnote\relax\footnote{This paper was presented in part at the IEEE International Symposium on Information Theory, Cambridge, Massachusetts, USA, July 2012 and the International Symposium on Turbo Codes and Iterative Information Processing, Gothenburg, Sweden, August 2012.}

\section{Introduction}
The multi-input multi-output (MIMO) systems are the transmission systems 
that use multiple antennas at both sides of the communication ends \cite{mimo1,mimo2}.
In the past few years, more and more attentions have been devoted to large MIMO systems, 
e.g., MIMO systems with tens to hundreds of transmit/receive antennas \cite{Large_MIMO_Scaling,Krylov,large_mimo_detector}
since the large MIMO systems 
can theoretically support very high data rate over wireless channels
without the need of  additional bandwidth and transmitted power.
Thus, the large MIMO systems which comprise of many low power antennas
could be a future trend for wireless communications that now have very scarce bandwidth.

Concatenating the MIMO systems with channel codes is the way
to considerably increase the reliability and the performance of the MIMO systems.
There has been a tremendous effort to develop the coded MIMO systems, 
including code-design, the invention of soft-output detectors, 
joint detection-decoding techniques and so forth, 
with the ultimate goal to approach the MIMO capacity, 
(a theoretical minimum SNR at which a reliable communication for the given rate is possible), 
see e.g. \cite{Hochwald_Turbo_MIMO,LDGM_MIMO,MCMC_MIMO}.
However, almost works have studied the coded MIMO systems
with small number of transmit/receive antennas (e.g. 2 to 8 transmit/receive antennas).
So, it is important to note that lacking in the literature is a performance study of coded large MIMO system.
We also note that the useful detection techniques for small MIMO systems, 
e.g., optimal MAP detection or  successive interference cancellation techniques, 
are not applicable to MIMO systems with hundreds of antennas since those methods 
definitely incur ultra high computational complexity at the receiver.

To the best of our knowledge,
only large MIMO systems concatenated with turbo codes have been studied in \cite{coded_large_mimo1,coded_large_mimo3,coded_large_mimo2,coded_large_mimo4}
which are the limited number of publications.
It is also worth mentioning that 
the main objective of the above cited works is on the development of low complexity large MIMO detections
not for an efficient coding scheme.
So, it is not surprising to see the near optimal detection performance (uncoded performance) but rather poor coded performance.
For example, turbo coded MIMO system with 200 transmit/receive antennas 
performs 7 dB away from the associated capacity 
which is quite large capacity-gap from the coding theoretic point of view.
One objective of this paper is to reduce such large capacity-gap
by considering the regular NBLDPC codes defined over $\GF(2^8)$ \cite{nb_ldpc} 
together with the low complexity detections.
The reasons why we consider the NBLDPC codes will be elaborated in the next section.

In this paper, we study the application of
NBLDPC codes to the MIMO systems with hundreds of antennas that utilize low complexity detection algorithms as the MIMO detector.
Our main contributions in this paper can be summarized as follows :

1) The NBLDPC coded large MIMO systems with well-known low-compelxity MMSE detector have been investigated.
By using the soft output generation method proposed in \cite{Wang_Soft_MMSE}, 
the NBLPDC coded MIMO systems
can operate nearer to the MIMO capacity comparing with the best known turbo coded systems \cite{coded_large_mimo1}.
For example, the capacity-gap about 15 dB for turbo coded large MIMO system with 16QAM can be reduced to 6 dB. 
Moreover, under the same soft-output MMSE detection, we have demonstrated via simulation that
NBLDPC codes significantly outperform both the optimized binary LDPC codes and regular binary LDPC codes over large MIMO systems.

2) For large MIMO systems, the potential drawback of using MMSE detection is the large size matrix inversion.
We thus propose the soft output detection based on matched filtering
which has lower computational complexity (just 0.28\% of MMSE detection).
Moreover, the similar coded performance, comparing to the case of using MMSE detection, 
in low SNR region or, equivalently, near capacity region can be obtained.

3) By using the Monte Carlo density evolution \cite{Davey_Thesis}
for analysing the NBLDPC coded large MIMO systems,
we found that the decoding threshold (minimum SNR at which the decoder can successfully decode the noisy received signals) of NBLDPC code 
is within 1.6 dB from the MIMO capacity.

4) Channel estimation is one of the challenges in large MIMO systems.
In this study, the sensitivity of the NBLDPC coded large MIMO systems to channel estimation error has been investigated.
For both the MMSE detection and the proposed detection,
large estimation error variance up to 0.2 is still acceptable for NBLDPC coded MIMO systems 
since the loss in coding gain is quite small (only 0.45 dB). 

5) Another one of the main concerns of large MIMO systems is how to place a large number of antennas in the limited space.
If the antenna spacing is not sufficient, e.g. spacing less than half of wavelength, 
the MIMO systems will suffer from spatial correlated fading
which results in the degradation of capacity as well as the coded performance.
In doubly correlated MIMO channel (correlation at both the transmitter and the receiver),
we observed that there exist the points in low SNR region at which the effect of spatial correlation is reduced.
At this region, the NBLDPC coded MIMO system with the proposed detection can be efficiently utilized.

The rest of this paper is organized as follows. 
We first describe the NBLDPC coded large MIMO systems in Section II.
Then the detection problem of large MIMO system is discussed in Section III. 
In Section IV, NBLDPC coded large MIMO systems with MMSE detection are described and investigated.
In Section V, we propose a new soft output detection for NBLDPC coded large MIMO systems.
In Section VI, the thresholds of NBLDPC coded large MIMO systems are analysed by Monte Carlo density evolution.
The sensitivity of NBLDPC coded large MIMO systems is studied in Section VII.
In Section VIII, the performance of NBLDPC coded large MIMO systems in correlated environment has been investigated.
Finally, conclusions are given in Section IX.

\section{System Description}
\subsection{Non-Binary Low-Density Parity-Check Codes}
An NBLDPC code $\mathrm{C}$ over Galois field $\GF(2^m)$ \cite{nb_ldpc}  is defined by 
the null-space of a sparse $P \times N$ parity-check matrix $\mathbf{A}=\lbrace a_{ij} \rbrace$ 
defined over $\GF(2^m)$, for $i=1,\ldots,P$ and $j=1,\ldots,N$
\begin{align*}
\mathrm{C} = \{ \mathbf{x} \in \GF(2^m)^{N} \mid \mathbf{A}\mathbf{x}^\mathsf{T} = \mathbf{0} \in \GF(2^m)^{P}\},
\end{align*}
where $m>1$ and $\mathbf{x} = (x_1,\ldots,x_N)$ is a codeword.
The parameter $N$ is the codeword length in symbol.
Assuming that $\mathbf{A}$ is of full rank, 
the number of information symbols is $K=N-P$ and the code rate is $R = K/N$.

We note that a non-binary symbol which belongs to $\GF(2^m)$ 
can be represented by the binary sequence of length $m$ bits.
For each $m$, we fix a $\GF(2^m)$ with a primitive element $\alpha$ and its primitive polynomial $\pi$. 
Once a primitive element $\alpha$ of $\GF(2^m)$ is fixed, 
each non-binary symbol is given by an $m$-bits representation \cite[p.~110]{macwilliams77}.
For example, with a primitive element $\alpha\in\GF(2^3)$ such that $\pi(\alpha)=\alpha^3+\alpha+1=0$, each symbol is represented as
$0=(0,0,0)$, $1=(1,0,0)$, $\alpha=(0,1,0)$, $\alpha^2=(0,0,1)$,
$\alpha^3=(1,1,0)$, $\alpha^4=(0,1,1)$, $\alpha^5=(1,1,1)$ and $\alpha^6=(1,0,1)$.
Let $L(x)$ be the binary representation of $x \in \GF(2^m)$.
For the above example, we can write $L(x=\alpha^3) = (1,1,0)$.
Thus, each coded symbol $x_i \in \GF(2^m), \forall i \in \lbrace1,\ldots,N\rbrace$ 
of a non-binary codeword represents $m$ bits.
We also denote $\mathrm{n} = mN$ and $\mathrm{k} = mK$ as the codeword length and information length in bit, respectively.

An NBLDPC code is $(\dv,\dc)$-regular if 
the parity-check matrix of the code has constant column weight $\dv$ and row weight $\dc$.
The parity-check matrix $\mathbf{A}$ can be represented by 
a Tanner graph with variable and check nodes \cite[p.~75]{mct_book}.
The belief propagation (BP) algorithm  which is commonly used as the NBLDPC decoder \cite{nb_ldpc} exchanges the probability vector
of length $2^m$ between variable nodes and check nodes of the Tanner graph  
at each iteration round $\ell$.

In this paper, only is $(\dv=2,\dc)$-regular NBLDPC code defined over $\GF(2^8)$ considered due to the following reasons :

1) The process to optimize parity-check matrix $\mathbf{A}$ is not required 
since $(2,\dc)$-regular NBLDPC code defined over $\GF(2^8)$
is empirically known as the best performing code especially for short code length.
Moreover, the NBLDPC code with $\dv=2$ can be encoded in linear time \cite{Lin_Time_NB_encoding}.

2) The high decoding complexity of NBLDPC decoder can be compensated
since the $\mathbf{A}$ of $(2,\dc)$-regular NBLDPC code is very sparse.

3) For practical code lengths, e.g., a few thousand bits, 
$(2,\dc)$-regular NBLDPC code defined over $\GF(2^8)$ seems to offer 
the best performance for small MIMO systems with optimal detection \cite{NB_MIMO_Peng}.
We therefore expect the excellent performance of NBLDPC codes in large MIMO systems.

4) It is empirically shown that the application of $(2,\dc)$-regular NBLDPC code defined over $\GF(2^8)$ 
to higher order modulation is outstanding 
\cite[p.~32]{ECC_4G_book}.

All good points listed above are very attractive for wireless communications.
Thus, it is not overly exaggerated to state that we intend to apply 
the simple, low-complexity, high-performance channel code to large MIMO systems.

\subsection{System Model}
We adopt the conventional notation 
to denote the MIMO system with $N_t$ transmit antennas and $N_r$ receive antennas
as $N_t \times N_r$ MIMO system.
Let $\mathbb{A}^{M}$ be the complex modulation constellation
of size $M=2^p$ where $p$ represents bit(s) per modulated symbol.
In this study, each antenna uses the same modulation scheme and the mapping is a Gray-labelled constellation.
We focus our attention to the spatial multiplexing MIMO systems \cite{sm_mimo1,sm_mimo2}
that simultaneously send $N_t$ modulated symbols in one time instant (one channel use). 
At the receiver side, some specific techniques will be employed to demultiplex the received data.

\begin{figure}[htb]
\centering
\includegraphics[scale=0.6]{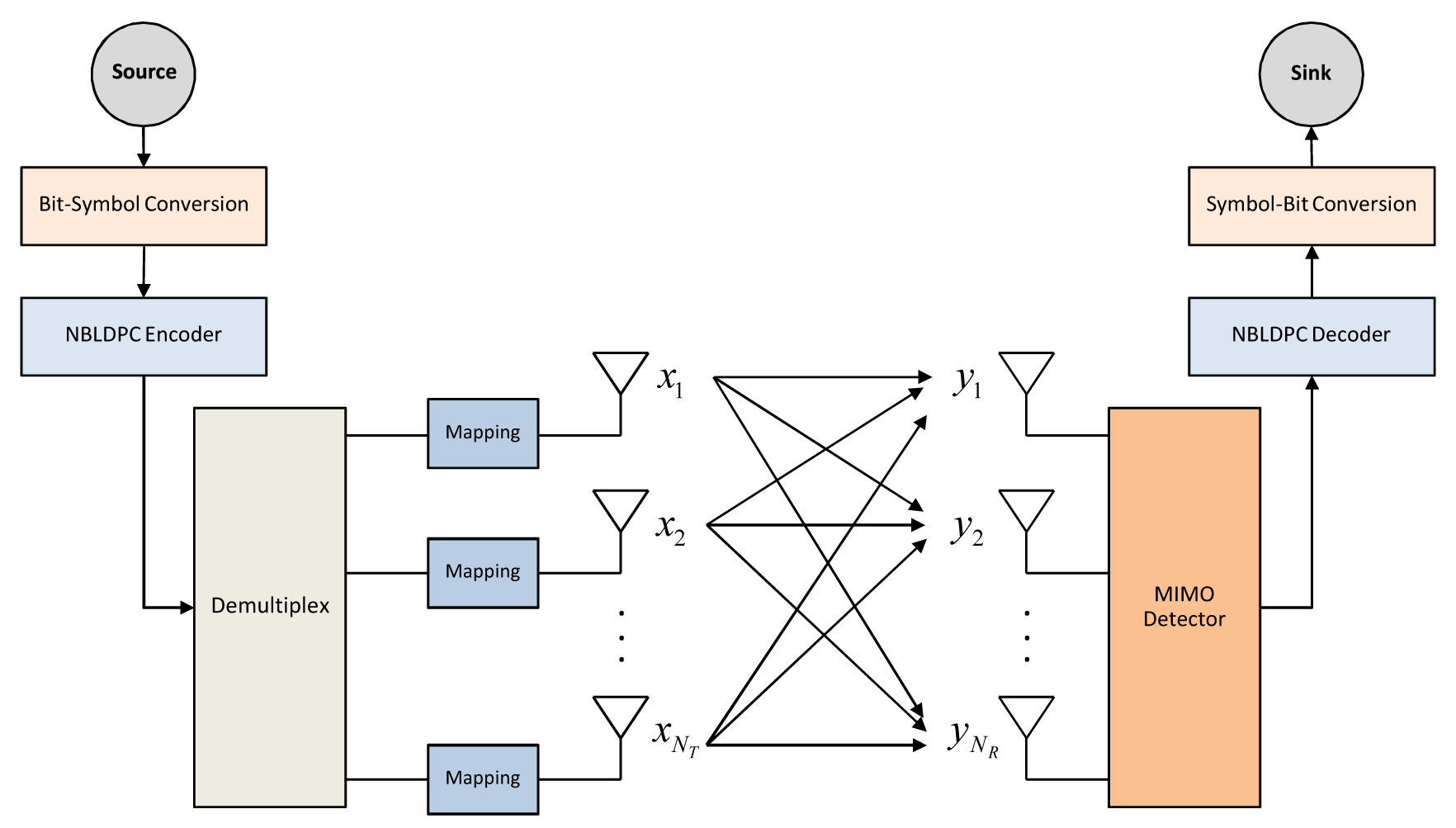}
\caption{Schematic diagram of the NBLDPC coded MIMO system. 
For NBLDPC code defined over $\GF(2^8)$,  every 8 bits emitted from source will be mapped to a symbol in $\GF(2^8)$.}
\label{NBLDPC_coded_MIMO}
\end{figure}

Figure \ref{NBLDPC_coded_MIMO} shows the spatial multiplexing MIMO system 
concatenated with an NBLDPC code.
In the case of NBLDPC code defined over $\GF(2^8)$, the system shown in Fig. \ref{NBLDPC_coded_MIMO} can be described as follows.
At the transmitter side, a bit to symbol mapper maps a group of 8 information bits 
to a non-binary symbol in $\GF(2^8)$.
The stream of $K$ symbols in $\GF(2^8)$ is encoded into a codeword of length $N$ symbols in $\GF(2^8)$ through an NBLDPC encoder.
Assuming $q \geq p$ and $q$ is divisible by $p$,
each coded symbol in $\GF(2^8)$ is demultiplexed to a group of $q=$ 8$/p$ symbols according to the chosen modulation scheme.
This means a coded symbol in $\GF(2^8)$ will be mapped to $q=$ 8$/p$ modulated symbols by a constellation mapping.
At each time instant, 
the transmitter simultaneously sends $N_t = K_tq$ modulated symbols in parallel 
through $N_t$ transmit antennas where $K_t$ is a number of coded symbols per each transmission.

Let $\mathbf{s} = [s_1,s_2,\ldots,s_{N_t}]^{T}\in\mathbb{C}^{N_{t}}$ be the transmitted vector whose each entry is a complex modulated symbol.
Each entry $s_i , \forall i \in \lbrace1,\ldots,N_t\rbrace $ taken from $\mathbb{A}^M$ must satisfy the component-wise power constraint $\mathrm{E}[\|s_i\|^2]= E_s/N_{t}$ 
where $E_s$ is the total transmitted power,  $\parallel\cdot\parallel^2$ denotes the squared Euclidean norm, and  $\mathrm{E}[ \cdot ]$ denotes the expectation.
With this power constraint, a large number of transmit antennas 
imply low power consumption per each transmit antenna.

Let us consider a $200 \times 200$ MIMO system with QPSK modulation $(M=4 \text{ and } p=2)$ as an example.
After encoding, each coded symbol in $\GF(2^8)$ is mapped to $q=4$ modulated symbols.
At each time instant, the transmitter collects 200 modulated symbols mapping from $K_t=50$ coded symbols.
These 200 modulated symbols are simultaneously sent through 200 transmit antennas in parallel. 

Considering a baseband discrete time model for an uncorrelated Rayleigh flat fading MIMO channel,
the received vector $\mathbf{y} = [y_1,y_2,\ldots,y_{N_r}]^{T}\in\mathbb{C}^{N_{r}}$ 
of the spatial multiplexing $N_t \times N_r$ MIMO system is given by \cite{coded_large_mimo1}
\begin{equation}
\label{model_eq}
\mathbf{y} = \mathbf{H}\mathbf{s} + \mathbf{n}.
\end{equation}
The matrix 
$\mathbf{H} = [\mathbf{H}_1 \mathbf{H}_2 \ldots \mathbf{H}_{N_t}] \in\mathbb{C}^{N_{r} \times N_{t}} $ 
denotes the channel fading matrix 
whose each entry $h_{kj}$ is assumed to be an i.i.d. complex Gaussian random variable 
with zero mean and unit variance $\mathrm{E}[\|h_{kj}\|^2]=1$.
The vector $\mathbf{n} = [n_1,n_2,\ldots,n_{N_r}]^{T}\in\mathbb{C}^{N_{r}}$ is a noise vector 
whose entry is an i.i.d. complex Gaussian random variable with zero mean and variance $\sigma^{2}_{n}$ per real component.
The MIMO detector performs detection and produces the prior probabilities (soft output) for NBLDPC decoder.
After all $N$ variable nodes are initialized, the NBLDPC decoder performs decoding process
and provides the estimated non-binary symbols (hard output).
These estimated symbols are finally demapped to a sequence of estimated information bits.

\subsection{SNR definition and Ergodic Capacity}

In this paper, the channel matrix $\mathbf{H}$ is assumed to be known at the receiver 
and we only focus on the square channel matrix, i.e. $N_t = N_r$.
Since each entry of $\mathbf{H}$ has unit variance, 
the average signal energy per receive antenna is $E_s$.
We follow the convention that $N_0/2 = \sigma^{2}_{n}$ to define the signal to noise ratio.
In this setting, the average signal to noise ratio (SNR) per receive antenna, 
denoted by $\gamma$, is given by \cite{Hochwald_Turbo_MIMO}
\begin{equation}
\label{SNR_eq}
\gamma = \frac{E_s}{N_0} = \frac{E_s}{2\sigma^{2}_{n}}.
\end{equation}

The spectral efficiency (transmitted information rate) of coded MIMO system with spatial multiplexing technique 
is $pRN_t$ \cite{Hochwald_Turbo_MIMO}.
With perfect $\mathbf{H}$ at the receiver side, ergodic MIMO capacity is given by \cite{mimo2}
\begin{equation}
\label{C_eq}
C = \mathrm{E}\left[ \log_2\det\left(\mathbf{I}_{N_r}+ \left(\gamma/N_t\right)\mathbf{H}\mathbf{H}^\mathsf{H}\right) \right],
\end{equation}
where the superscript $\mathsf{H}$ denotes the Hermitian transpose operator, $\det$ denotes the determinant 
and $\mathbf{I}_{N_r}$ is the identity matrix of size $N_r \times N_r$.
Both of spectral efficiency and capacity are measured in bits/sec/Hz (bps/Hz).

\subsection{Large MIMO Detection}
One of the major problems 
that prohibits large MIMO systems from the practical realization
is the high computational complexity of the MIMO detector.
To clearly show this, let us illustrate by using an example.
Suppose that the number of transmit antennas are $N_t=100$ and the BPSK modulation is employed.
A total number of possible sequences that can be transmitted from the MIMO transmitter 
is $2^{pN_t} = 2^{100}$ sequences which are very large number.
Therefore, performing optimal maximum likelihood detection, 
i.e., finding the most likely transmitted vector given the received signal, 
on large MIMO systems is infeasible.
We also note that the near optimal detection algorithms inventing for small MIMO systems, 
e.g., sphere detection or successive interference cancellation, 
are also too complex for large MIMO systems.
Therefore, many works have recently focused on inventing the low-complexity detectors 
to enable the use of large scale multiple antennas \cite{large_mimo_detector,coded_large_mimo1,coded_large_mimo3,coded_large_mimo4}.

In order to operate near capacity region over large MIMO systems, we exactly need the efficient coding and detection scheme.
Good detection performance (uncoded performance), e.g., near optimal detection performance, does not imply
good coded performance, e.g., near capacity performance.
For example, likelihood ascent search (LAS) detection can provide near optimal detection performance 
but the capacity-gap from using Turbo codes is more than 15 dB for $600 \times 600$ MIMO systems with 16-QAM \cite{coded_large_mimo1}.
This is because, for modern channel codes such as  Turbo and LDPC codes, 
good coded performance needs the generation of soft output from the MIMO detector as addressed in \cite{coded_large_mimo3}.
Thus, we believe that the generation of soft output from the MIMO detector is very crucial 
to obtain  the good coded performance from using the NBLDPC codes.

\begin{figure}[htb]
\centering
\includegraphics[scale=0.7]{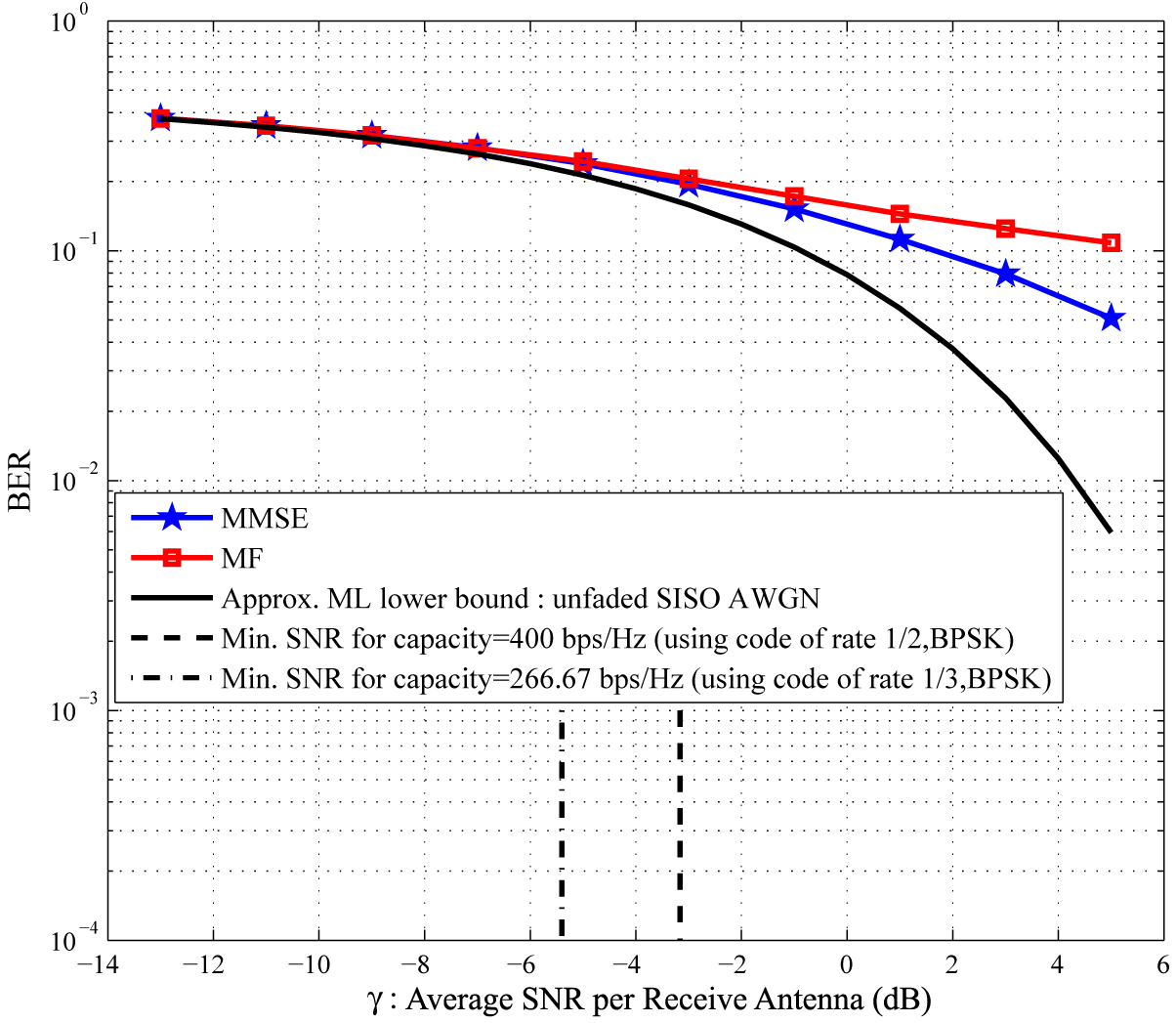}
\caption{Uncoded performance of MMSE and MF detectors in $800\times800$ MIMO system with BPSK modulation.}
\label{uncoded_BER}
\end{figure}

For large MIMO systems,
it is well-known that two types of low-complexity linear detections, 
named as minimum mean square error (MMSE) and the matched filter (MF) detections,
can provide the near optimal detecting performance especially in low SNR region \cite{Large_MIMO_Scaling}.
The results given in
Fig. \ref{uncoded_BER} which displays the detecting performance of MMSE and MF detections over $800 \times 800$ MIMO systems with BPSK modulation can be used to confirm the above statement.
In this figure, the uncoded performance of SISO unfaded AWGN-BPSK system is used
to represent the approximated lower bound for optimal maximum likelihood (ML) performance \cite{tabu_large_mimo}.
From this figure, the major observations could be made :

$\bullet$ There is no performance difference between MF and MMSE detections when  $\gamma$ is low, e.g.,  $\gamma \leq -2$ dB.

$\bullet$ The performance of MMSE and MF detections at low SNRs achieves near the performance of SISO unfaded AWGN.
Thus, both are near optimal in low SNR region.

$\bullet$ The minimum SNR to achieve capacity below 400 bps/Hz is in the near optimal region 
of both the MMSE and MF detections.
So, it is intuitive to think that 
the MMSE and MF detections with soft output generation can be used in conjunction with 
the NBLDPC codes that have low to medium rates
to achieve near capacity performance.

Motivated by the above discussion, we will focus in this work the MMSE and MF detections.
An important thing that need to be subsequently pointed out is 
the methodology to obtain the soft output from MIMO detector.
In the next section, we will present the detection principle of MMSE detection as well as the suitable soft output generation that works well with NBLDPC decoder.
In Section V, we will develop the novel soft output MF-based detection that can provide excellent coded performance 
with very low computational complexity.

\section{MMSE Detection}
Let us first introduce in this section the MMSE detector and  its soft output generation.
After that, the coded performance will be investigated and discussed.

\subsection{Detection Principle}
Following \cite{soft_mmse}, 
a detection estimate $\hat{s}_k$ of the transmitted symbol on $k$-th antenna
by multiplying $y_k$ with the MMSE weight vector $\mathbf{W}_k$
\begin{equation}
\label{MMSE_detect_eq}
\hat{s}_k = \mathbf{W}_{k}^\mathsf{H} y_k,
\end{equation}
where the MMSE weight matrix $\mathbf{W}_k$ is of the form
\begin{equation}
\label{MMSE_eq}
\mathbf{W}_{k} = \left( \frac{N_0}{E_s/N_t} \mathbf{I}_{N_r} + \mathbf{H}\mathbf{H}^\mathsf{H}  \right)^{-1} \mathbf{H}_k.
\end{equation}
This MMSE weight vector $\mathbf{W}_k$ is chosen so as to minimize the mean square error 
between the transmitted symbol $s_k$ and $\hat{s}_k$.

\subsection{Soft Output Generation}
By using the method for obtaining soft output from MMSE detector given in \cite{Wang_Soft_MMSE}, the estimation $\hat{s}_k$ can be approximated as 
the output of an equivalent additive white Gaussian noise (AWGN) channel 
\begin{equation}
\label{Eq_AWGN_eq}
\hat{s}_k = \mu_k s_k + z_k,
\end{equation}
where $ \mu_k = \mathbf{W}_{k}^\mathsf{H} \mathbf{H}_k $ 
and $z_k$ is a zero-mean complex Gaussian variable 
with variance $\epsilon^{2}_{k} = \frac{E_s}{N_t} ( \mu_k - \mu_{k}^{2} )$.
Based on this approximation, 
the probability of $\hat{s}_k$ conditioned on $s \in \mathbb{A}^{M}$ is as follows
\begin{equation}
\label{Soft_MMSE}
\mathrm{Pr}\left( \hat{s}_k \mid s \right) \simeq  \frac{\kappa}{\pi\epsilon^{2}_{k}} \text{exp} \left( -\frac{1}{\epsilon^{2}_{k}} \parallel \hat{s}_k - \mu_k s \parallel^2 \right),
\end{equation}
where the parameter $\kappa$ is the normalized constant such that 
$\sum_{s \in \mathbb{A}^M } \mathrm{Pr}\left( \hat{s}_k \mid s \right) = 1 $.
For $m\neq p$ which is the condition to be considered, 
the probability $\mathrm{Pr}\left( \hat{s}_k \mid s \right)$ cannot feed directly to NBLDPC decoder. 

\begin{figure}[htb]
\centering
\includegraphics[scale=1.2]{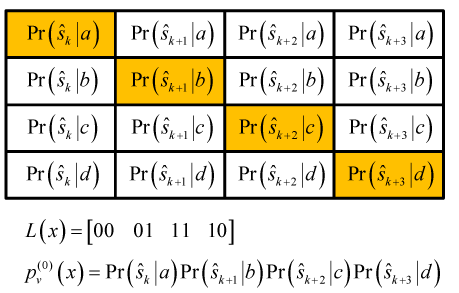}
\caption{The example of calculating soft output for symbol $x$ with $L(x)=\left[ 00~01~11~10 \right]$.
In this example, we assume that the complex modulation constellation of size $M$ = 4, e.g., QPSK, is employed
and the elements in this constellation are $a,b,c,d$ with their gray-labelled representations $00,01,11,10$, respectively.}
\label{Prob_NB_dec}
\end{figure}

By deploying the iterative belief propagation (BP) algorithm as an NBLDPC decoder \cite{BP_dec_GFq},
the procedure to form the input of NBLDPC decoder which is defined over $\GF(2^8)$ 
from the known probabilities calculating from (\ref{Soft_MMSE}) is as follows.
Let $p^{(0)}_{v}(x)$ denote the probability that $v$-th coded symbol is likely to be $x \in \GF(2^8)$
where $v = 1,2,\ldots,N$ and the index in superscript represents the iteration round of BP algorithm.
We assume that $k$-th transmit antenna to $(k+q-1)$-th transmit antenna 
are used to send the $v$-th coded symbol (any coded symbol is mapped to $q$ modulated symbols).
For each $x \in \GF(2^8)$, we need to collect $q$ modulated symbols to represent $qp = 8 $ bits.
Thus, the generation of soft output from MMSE detector for the symbol $x \in \GF(2^8)$ of
$v$-th variable node can be expressed as
\begin{equation}
\label{input_eq}
p^{(0)}_{v}(x) = \prod^{q-1}_{i=0}\mathrm{Pr}\left( \hat{s}_{ k+i } \mid s^{ k+i } \right),
\end{equation}
where $\mathrm{Pr}\left( \hat{s}_{ k+i } \mid s^{ k+i } \right)$ is the likelihood of the estimation $\hat{s}_{k+i}$ 
conditioned on $s^{k+i} \in \mathbb{A}^{M}$ 
by which the eight bits ordered sequence of $( s^{k}, s^{k+1}, \cdots s^{k+q-1} )$ 
must be equal to the binary representation $L(x)$. 
An example of calculating $p^{(0)}_{v}(x)$ is illustrated in Fig. \ref{Prob_NB_dec}.

According to (\ref{input_eq}), 
$2^8(q-1)$ real multiplications for each coded symbols are needed 
to calculate $p^{(0)}_{v}(x)$ for all $x \in \GF(2^8)$.
We note that the computational complexity of generating soft output for NBLDPC decoder is rather low
when comparing to the computation of MMSE matrix.

\subsection{Performance}
In this part, the performance of NBLDPC coded MIMO systems with soft output MMSE detector is presented.
The maximum iteration of NBLDPC decoder $\ell_{\textrm{max}}$ is set to 200 for all simulation results \cite{ISIT_LARGE}.

\begin{figure}[htb]
\centering
\includegraphics[scale=0.7]{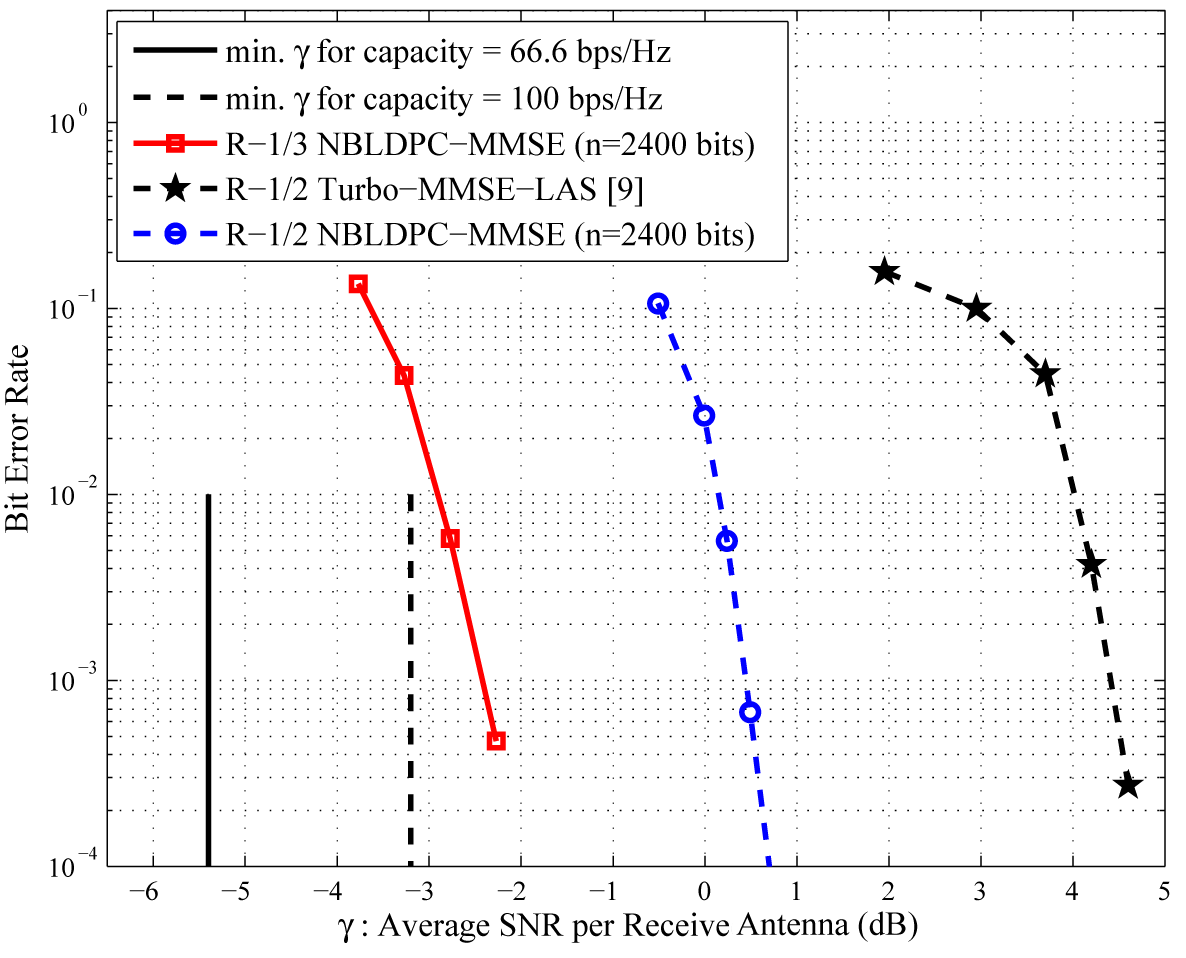}
\caption{Bit error rate performances of coded $200 \times 200$ MIMO systems with BPSK modulation.
The spectral efficiencies of 66.6 and 100 bps/Hz 
are obtained from MIMO systems with channel codes of $R=1/3$ and $R=1/2$ respectively.}
\label{ber200x200}
\end{figure}

The BER performance of NBLDPC coded $200 \times 200$ MIMO systems 
with BPSK modulation $(M=2)$ is shown in Fig. \ref{ber200x200}.
For $200 \times 200$ MIMO system,
the best performing scheme which can be founded in the literature is
the $R=1/2$ turbo coded spatial multiplexing MIMO systems with the MMSE-LAS detector \cite{coded_large_mimo1}. 
The MMSE-LAS detection algorithm uses the MMSE detection to initialize the algorithm.
Therefore, the overall computational complexity of MMSE-LAS detector is greater than that of MMSE detector.
The BER performance of this turbo coded system is away from MIMO capacity by 7.5 dB.
It is clearly seen from the figure that
$R=1/2$ NBLDPC coded system significantly outperforms turbo coded system by about 4 dB.
Both $R=1/3$ and $R=1/2$ NBLDPC coded systems perform within just 3.5 dB from the MIMO capacity. 

\begin{figure}[htb]
\centering
\includegraphics[scale=0.7]{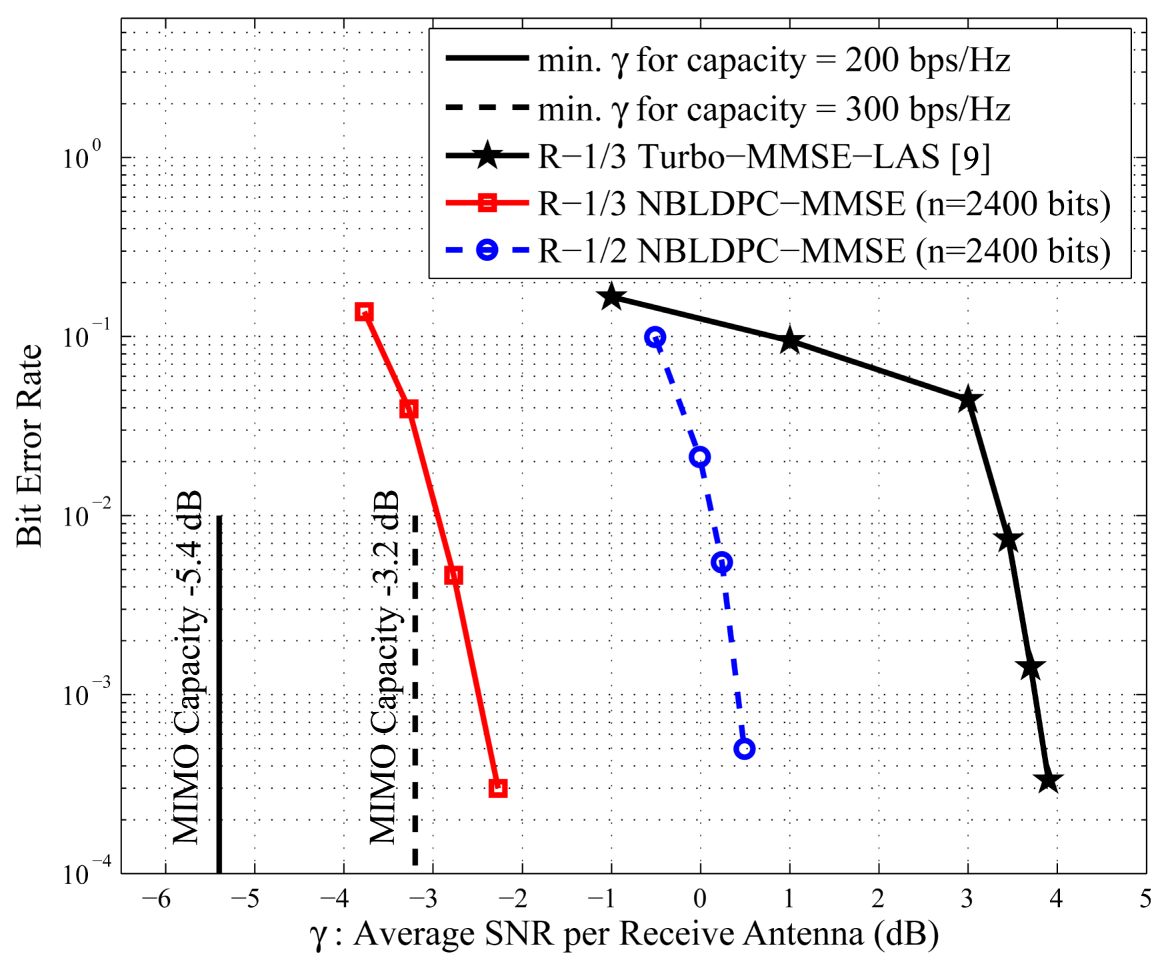}
\caption{Bit error rate performances of coded $600 \times 600$ MIMO systems with BPSK modulation.
The spectral efficiencies of 200 and 300 bps/Hz 
are obtained from MIMO systems with channel codes of $R=1/3$ and $R=1/2$ respectively.}
\label{ber600x600}
\end{figure}

Figure \ref{ber600x600} presents the simulation results for larger dimension, $600 \times 600$ MIMO system.
Comparing to $200 \times 200$ MIMO system,
it can be observed that the capacity-gaps remain the same (3.5 dB) for both $R=1/3$ and $R=1/2$ NBLDPC coded systems.
However, the capacity-gap of turbo coded system with MMSE-LAS detector in \cite{coded_large_mimo1}  is larger (about 9.4 dB) when 
the dimension of channel increases.
It is obviously seen that $R=1/3$ turbo coded system underperforms $R=1/3$ NBLDPC coded system by more than 6 dB.
More interestingly, $R=1/2$ NBLDPC coded system absolutely outperforms $R=1/3$ turbo coded system by about 3.4 dB.
Therefore, the NBLDPC coded system outperforms the turbo coded system both in terms of performance and spectral efficiency.

\begin{figure}[htb]
\centering
\includegraphics[scale=0.7]{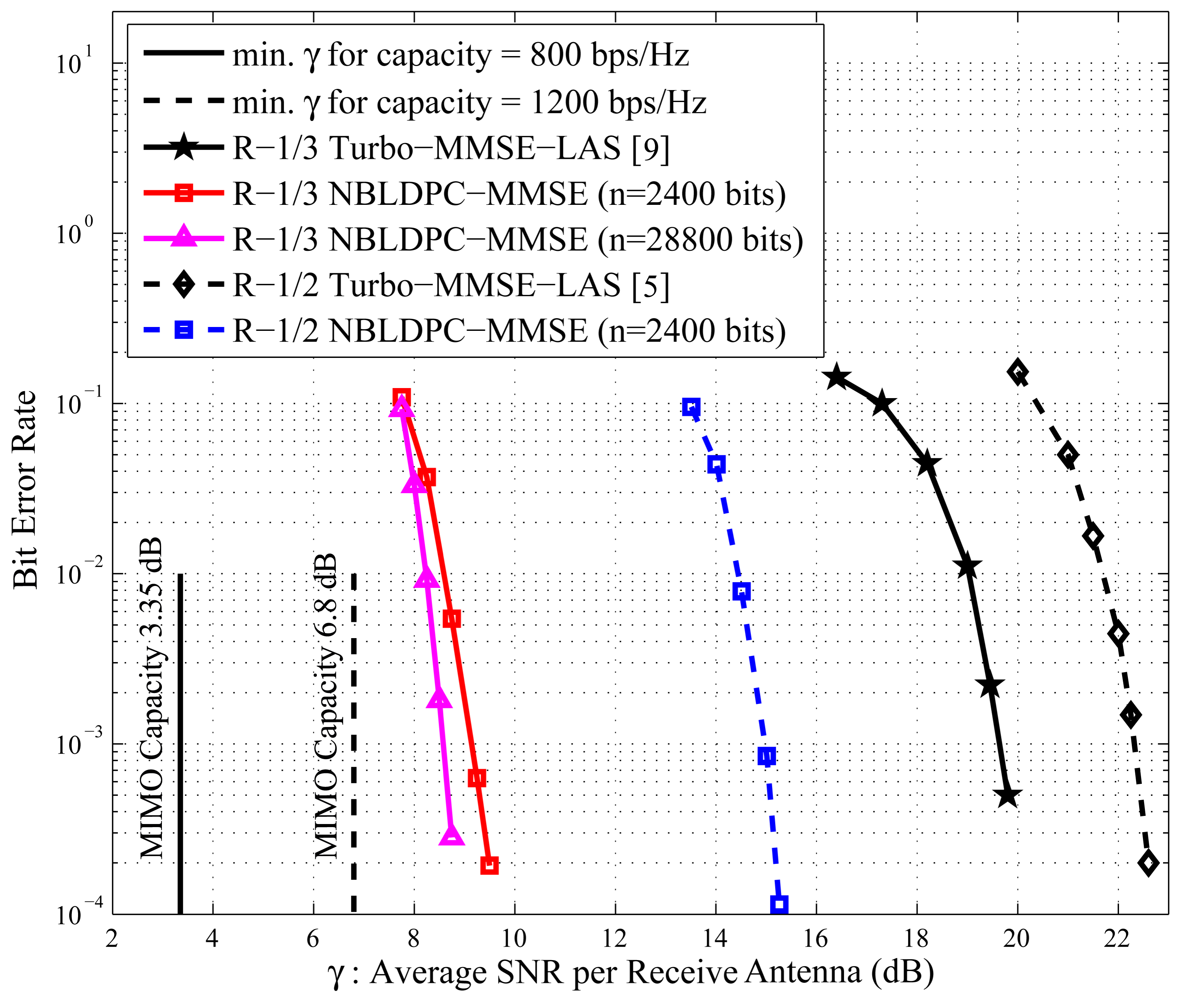}
\caption{Bit error rate performances of coded $600 \times 600$ MIMO systems with 16-QAM modulation.
The spectral efficiencies of 800 and 1200 bps/Hz 
are obtained from MIMO system with channel codes of $R=1/3$ and $R=1/2$ respectively.}
\label{ber600x600_16QAM}
\end{figure}

It is well known that the spectral efficiency can be enhanced 
without additional bandwidth by adopting higher order modulations 
such as $M$-QAM or $M$-PSK.
Thus, it is interesting to investigate the performance of coded large MIMO system with higher order modulation.
The coded performance of $600 \times 600$ MIMO system with $16$-QAM is 
shown in Fig. \ref{ber600x600_16QAM}.
From this figure, the following observations can be listed as follows :

\noindent$\bullet$ $R=1/3$ and $R=1/2$ NBLDPC coded systems with $\mathrm{n}=2400$ bits
operate within 6 dB and 8 dB, respectively, from MIMO capacity.

\noindent$\bullet$ $R=1/3$ and $R=1/2$ turbo coded systems \cite{coded_large_mimo1} 
operate very far (more than 15 dB) from MIMO capacity. 
It is clearly seen that the NBLDPC coded systems indeed beat turbo coded systems. 

\noindent$\bullet$ By increasing the length of code from $n = 2400$ bits to $n=28800$ bits, 
the coding gain about 0.7 dB can be obtained from $R=1/3$ NBLDPC coded system.

Based on the result, we can conclude that the good performance over large MIMO channels
by using NBLDPC codes and 16-QAM are possible.

One may think that the previous results are not sufficient to show the benefit of using NBLDPC code in large MIMO systems.
Moreover, we conduct the comparison among different types of LDPC codes : regular NBLDPC code over $\GF(2^8)$, optimized binary LDPC code, and regular binary LDPC code.
By fixing the code rate $R=1/3$, the regular binary LDPC code with column weight 4 and row weight 6
and the optimized binary LDPC code from Table. III in \cite{Hou_opt_bldpc} 
are chosen as the competitors.
The maximum degree of variable node of the selected optimized binary LDPC code is 16.
We choose this optimized LDPC code because it can asymptotically perform very close to 
the capacity of single input single output Rayleigh fading channel (within 0.19 dB).
By using the same 16-QAM and soft output MMSE detection,
the performances of $600 \times 600$ MIMO systems coded with different kinds of LDPC codes are 
shown in Fig. \ref{ber600x600_16QAM}.
From this figure, the interesting observation can be listed as follows :

\noindent$\bullet$ $R=1/3$ NBLDPC coded systems outperform both the optimized and regular binary LDPC (BLDPC) codes. 
The coding gains obtained over optimized and regular binary LDPC codes are about 0.8 and 2 dB respectively.

\noindent$\bullet$ Another advantage of using NBLDPC code which can be seen from this figure
is the excellent frame error rate (FER) which is an advantage that can be expected from LDPC codes of column weight two \cite{recursive_ldpc_wc2}.
Comparing with optimized BLDPC code, the FER of NBLDPC code is much better. 
Although the BER curve of optimized BLDPC code is good but its corresponding FER curve is quite bad.

\begin{figure}[htb]
\centering
\includegraphics[scale=0.7]{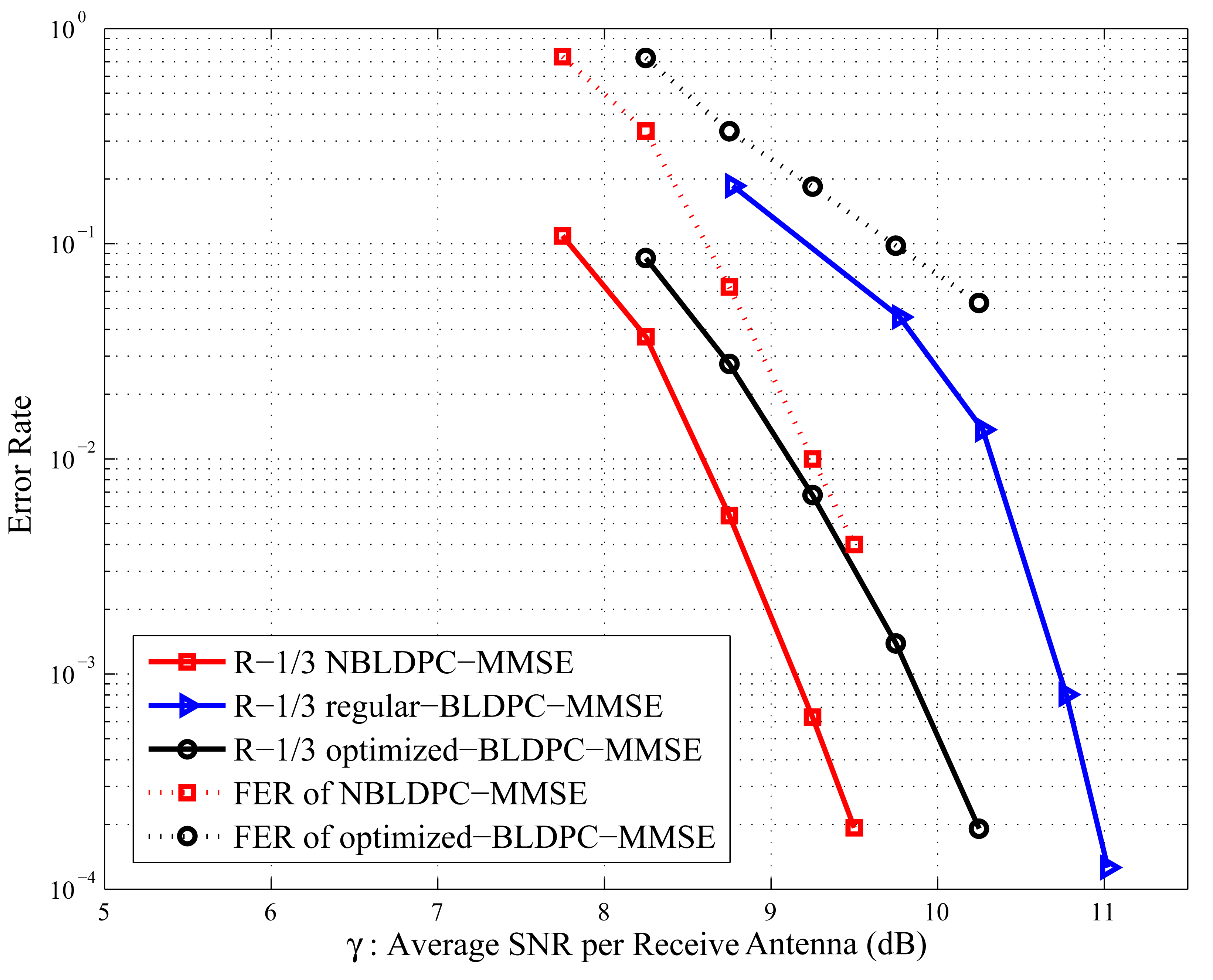}
\caption{Performance comparison between $R=1/3$ NBLPDC code and $R=1/3$ BLDPC codes on $600 \times 600$ MIMO systems with 16-QAM modulation at $\mathrm{n}=2400$ bits.
The solid curves represent the BER performance while the dashed curves are the corresponding FER performance.}
\label{comparison600x600_16QAM}
\end{figure}

We have shown in this section that NBLDPC coded large MIMO systems with soft output MMSE detection 
can be employed to reduce the remaining capacity-gap of the best known turbo coded systems.
Despite the low complexity of MMSE detection, one may think that finding inverse of large and dense channel matrix  which is the main task of MMSE detection is rather cumbersome.
Inspired by the very low-complexity and near optimal detecting performance of MF detector, 
the soft output MF-based detector will be developed in the next section.

\section{The Proposed MF-Based Detection}
\subsection{Detection Principle}
A baseband discrete time channel model given in (\ref{model_eq})
can be equivalently rewritten as 
\begin{equation}
\label{model2_eq}
\mathbf{y} = \mathbf{H}_1 s_1 + \mathbf{H}_2 s_2 + \cdots + \mathbf{H}_{N_t} s_{N_t} + \mathbf{n}.
\end{equation}
To estimate $\hat{s}_k$ of the transmitted symbol on the $k$-th antenna by
the concept of matched filtering,
the interference streams in (\ref{model2_eq}) are treated as noise,
\begin{equation}
\label{assump_eq}
\mathbf{y} = \mathbf{H}_k s_k + \underbrace{ \sum\limits_{i=1,i \neq k}^{N_t}\mathbf{H}_i s_i + \mathbf{n} }_{\textrm{noise}}.
\end{equation}
Then, the MIMO detector based on MF technique
estimates $\hat{s}_k$ of the transmitted symbol on the $k$-th antenna
by multiplying the received vector $\mathbf{y}$ with a vector
$\mathbf{W}_k =\frac{\mathbf{H}_k^\mathsf{H}}{\mathbf{H}_k^\mathsf{H}\mathbf{H}_k}$,
\begin{equation}
\label{detect_eq}
\begin{array}{ll}
   \hat{s}_k&= \mathbf{W}_k\mathbf{y},\\
&= s_k + \mathbf{W}_k\sum\limits_{i=1,i \neq k}^{N_t}\mathbf{H}_i s_i + \mathbf{W}_k\mathbf{n}.
\end{array}
\end{equation}

In contrast of the conventional MF detection, 
we make a little modification by introducing the term $\frac{1}{\mathbf{H}_k^\mathsf{H}\mathbf{H}_k}$ into weight vector.
After slicing $\hat{s}_k$,
we can obtain the estimated symbol transmitted from the $k$th antenna.
It is known that the detection performance (uncoded performance) of match filtering is very poor 
since there is no cancellation of the interference terms (the second term in R.H.S of (\ref{detect_eq})).
However, matched filtering is near optimum if $\mathbf{n}$ is dominant \cite{Large_MIMO_Scaling,massive_mimo}, i.e., at low SNRs, as shown in Fig. \ref{uncoded_BER}. 

\begin{figure}[htb]
\centering
\includegraphics[scale=0.75]{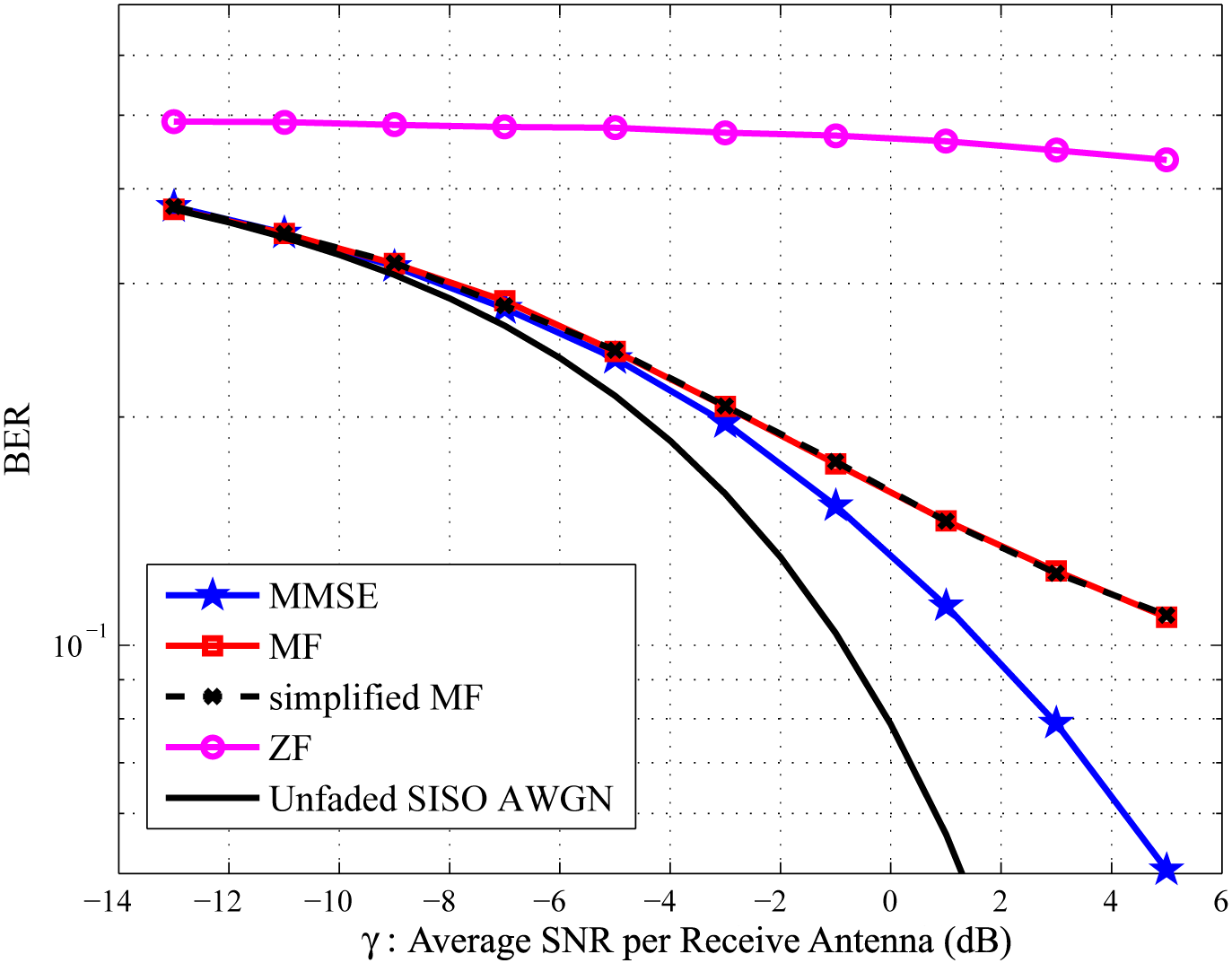}
\caption{Uncoded performance of MMSE, ZF, MF, and simplified MF detectors in $200\times200$ MIMO system with BPSK modulation.
The uncoded performance of SISO unfaded AWGN-BPSK system is again used
to represent the approximated lower bound for ML performance \cite{tabu_large_mimo}.
The performance of zero forcing (ZF) detection which is a well-known low-complexity detector is also shown.
The ZF detection performs very poor in low SNRs
so we do not pay the attention to this type of detector.}
\label{MF_uncode}
\end{figure}

To reduce the computational complexity of detection,
we observe that $\mathrm{E}\left[\mathbf{H}_k^\mathsf{H}\mathbf{H}_k\right] \approx N_r$.
Thus, in the case of hundreds of antennas, the computation of $\mathbf{W}_k$ can be simplified as follows
\begin{equation}
\label{reduc1}
\begin{array}{ll}
\mathbf{W}_k & = \mathbf{H}_k^\mathsf{H} \left( 1/N_r \right) .\\
\end{array}
\end{equation}
With this simplification,
the matrix multiplication $\mathbf{H}_k^\mathsf{H}\mathbf{H}_k$ whose an complexity is $\mathcal{O}(N_r)$ \cite{matrix_computation} can be reduced to just a constant term.
Figure \ref{MF_uncode} shows that there is no performance difference between MF detection and its corresponding simplified version.

\subsection{Soft Output Generation}
In this part, a novel soft output generation from MF-based detection for NBLDPC decoder 
which has not yet reported elsewhere is presented.
Assuming all streams and noise are statistical independence and Gaussian distributed,
the proposed method for generating soft output from MF-based detection will be described as follows.
Following the definition of the post detection signal to interference plus noise ratio ($\mathrm{SINR}$) given in \cite[p.~358]{Cho_MIMO_book}, 
the $\mathrm{SINR}$ of the proposed detection for $k$-th stream, denoted by $\delta_k$, can be expressed as
\begin{equation}
\label{SINR_eq}
\begin{array}{ll}
   \delta_k& = \frac{\textbf{$\mathrm{E}[\Vert s_k \Vert^2]$}}
   {\textbf{$\mathrm{E}[\sum\limits_{i=1,i \neq k}^{N_t} \Vert \mathbf{W}_k\mathbf{H}_i s_i \Vert^2]$}
   ~+~\textbf{$\mathrm{E}[\Vert \mathbf{W}_k\mathbf{n} \Vert^2]$}},\\
 & = \frac{E_s/N_t}{E_s/N_t\sum\limits_{i=1,i \neq k}^{N_t}\|\mathbf{W}_k\mathbf{H}_i\|^{2}~+~\left(2\sigma^{2}_{n}\|\mathbf{W}_k\|^{2}\right)},\\
 & = \frac{E_s/N_t}{\Delta_k}. \\
\end{array}
\end{equation}
The denominator, denoted by $\Delta_k$, given in (\ref{SINR_eq}) can be approximated by a Gaussian random variable.
Based on this approximation, 
we set $\sigma^{2}_{k} = \Delta_k/2$ 
and the soft output which exactly is the likelihood of $\hat{s}_k$ conditioned on $s \in \mathbb{A}^{M}$ is as follows
\begin{equation}
\label{soft_MF}
\mathrm{P}\left( \hat{s}_k \mid s \right) =  \frac{1}{\sqrt{2\pi\sigma^{2}_{k}}}~ \text{exp} \left( -\frac{1}{2\sigma^{2}_{k}}\|\hat{s}_k - s\|^{2}\right).
\end{equation}

\begin{figure}[htb]
\centering
\includegraphics[scale=0.75]{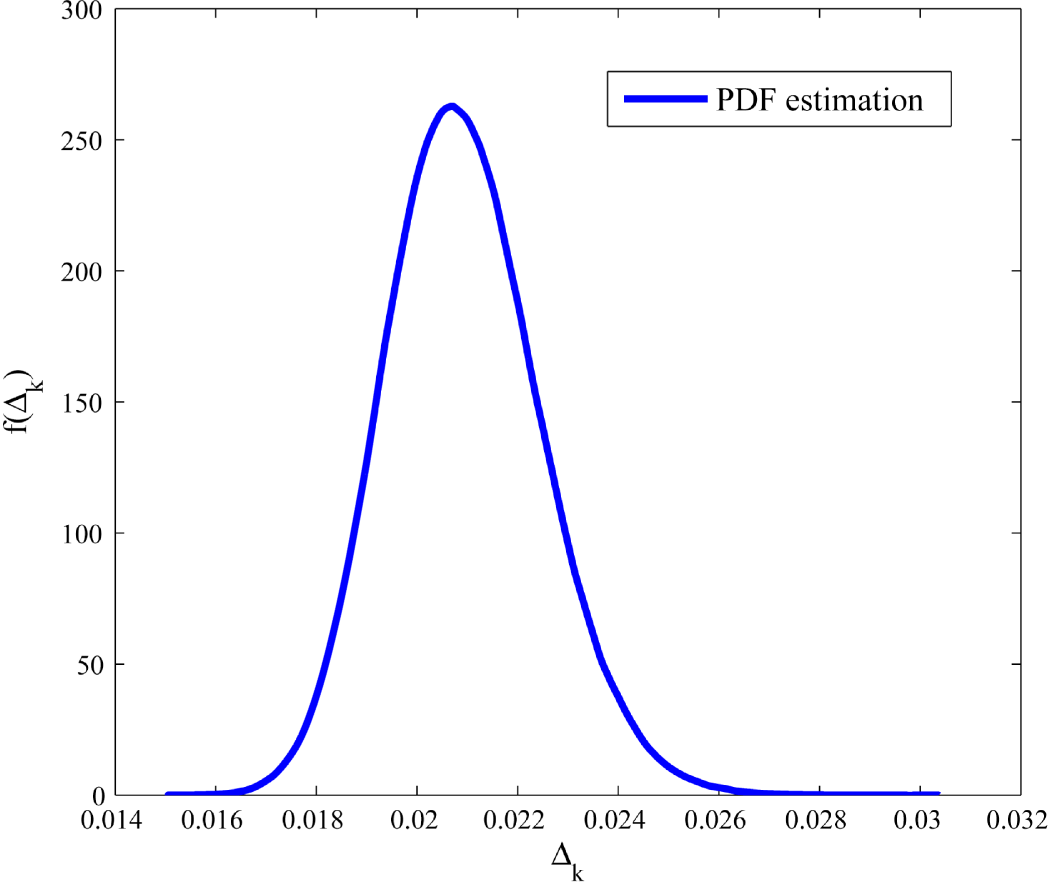}
\caption{The $\mathrm{pdf}$ estimation of $\Delta_k$ from $200 \times 200$ MIMO channel with BPSK modulation at SNR per receive antenna $\gamma$ = -2 dB.}
\label{PDF_estimation}
\end{figure}

To justify that $\Delta_k$ distributes like Gaussian,
we estimate the probability density function ($\mathrm{pdf}$) of random variable $\Delta_k$ 
by mean of Kernel density estimation \cite{prob_book}.
By using 100,000 channel realizations, 
it is obviously seen from Fig. \ref{PDF_estimation} 
that the $\mathrm{pdf}$ of $\Delta_k$ closely agrees with the shape of Gaussian distribution.
Moreover,  we also compare the distribution of $\Delta_k$ and Gaussian distribution by means of Kolmogorov-Smirnov test \cite[pp.~623-626]{Numer_Book}.
It is found that both have the same distribution at the 0.1$\%$ significance level.
Therefore, we conclude that the Gaussian approximation of $\Delta_k$ described above is reasonable.
By using (\ref{input_eq}), the input for NBLDPC decoder can be now formalized from (\ref{soft_MF}).

To reduce the computational complexity of soft output generation,
we first observe that $\mathrm{E}\left[ \|\mathbf{W}_k\|^{2} \right] \approx 1/N_r$. 
This is because $\mathrm{E}\left[ \|\mathbf{H}_k^{\mathsf{H}} \|^{2} \right] \approx N_r$ 
and $\mathrm{E}\left[\|\mathbf{H}_k^\mathsf{H}\mathbf{H}_k\right \|^{2}] \approx N_r^2$.
Then, we assume that the power of noise term is dominant, i.e., $2\sigma^{2}_{n} \gg E_s$.
With this assumption, we have the following expression
\begin{align*}
E_s/N_t\sum\limits_{i=1,i \neq k}^{N_t}\|\mathbf{W}_k\mathbf{H}_i\|^{2} \ll  2\sigma^{2}_{n}\|\mathbf{W}_k\|^{2}.
\end{align*}
The above expression is valid when $\gamma$ is low, e.g., $\gamma$ = -5 dB,
since $1/N_t\sum\limits_{i=1,i \neq k}^{N_t}\|\mathbf{W}_k\mathbf{H}_i\|^{2} < \|\mathbf{W}_k\|^{2}$.
Therefore, the denominator term $\Delta_k$ in (\ref{SINR_eq}) can be simplified as follows,
\begin{equation}
\label{reduc2}
\begin{array}{ll}
\Delta_k & = E_s/N_t\sum\limits_{i=1,i \neq k}^{N_t}\|\mathbf{W}_k\mathbf{H}_i\|^{2}~+~\left(2\sigma^{2}_{n}\|\mathbf{W}_k\|^{2}\right),\\
 & \approx  2\sigma^{2}_{n}\|\mathbf{W}_k\|^{2},\\
 & \approx 2\sigma^2_n/N_r.
\end{array}
\end{equation}
With this simplification, the computation of variance $\sigma^2_k$ for soft output generation
defined in (\ref{Prob_eq}) can be reduced to a constant  term $\sigma^2_k= \Delta_k/2 = \sigma^2_n/N_r$.
The variance $\sigma^2_k$ for soft output generation is now independent of $k$ and can be easily pre-computed.
We note again that the method for generating soft output from MF-based detector presented in this part
is completely different from the previously found techniques for obtaining soft output from MF detection \cite{Ito_MF,scaled_matched_filter}.

\subsection{Performance}
\begin{figure}[htb]
\centering
\includegraphics[scale=0.7]{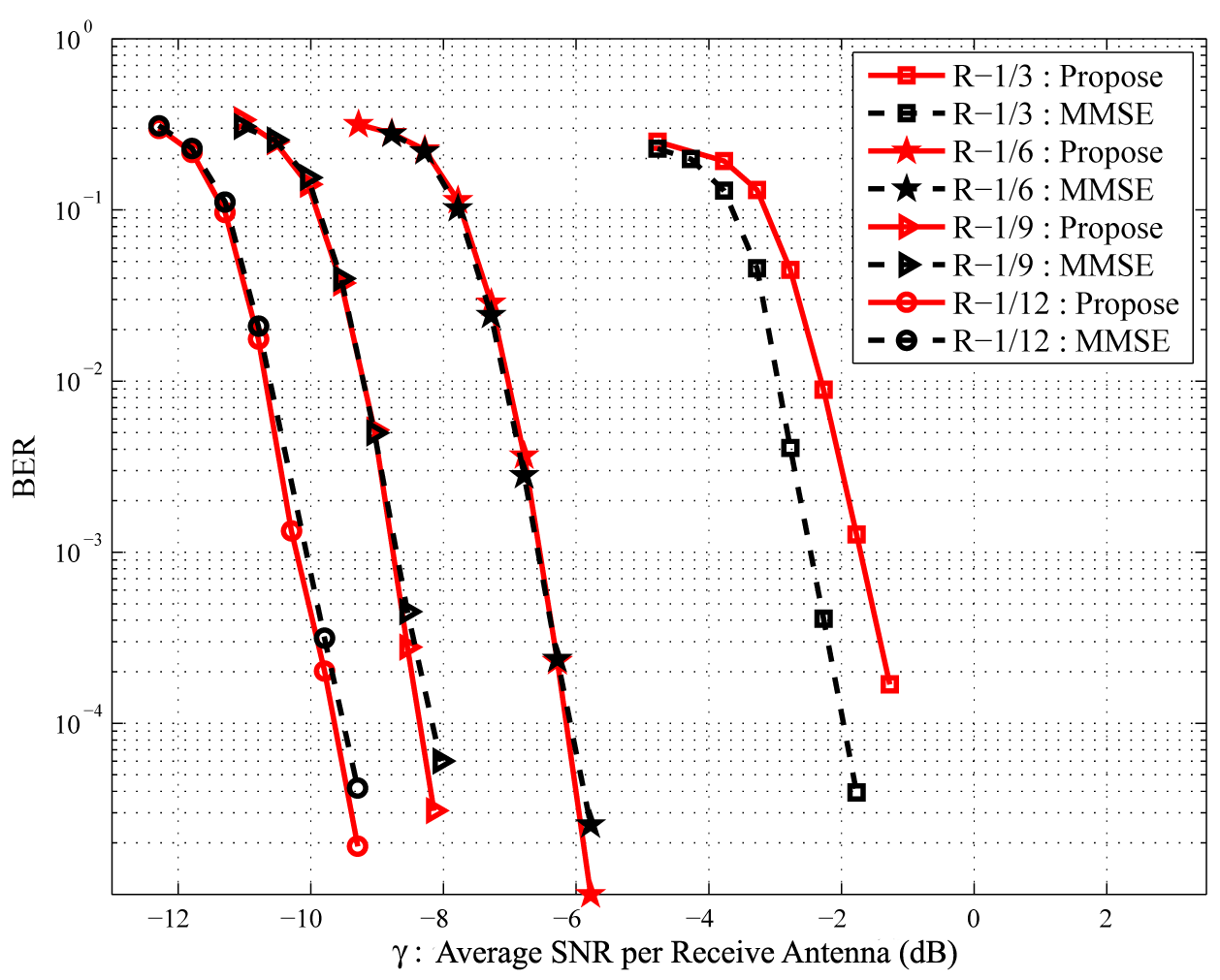}
\caption{BER of the NBLDPC coded $200 \times 200$ MIMO systems with BPSK modulation. 
``Propose" means the NBLDPC coded system with the proposed soft output MF-based detection and ``MMSE" represents the non-binary LDPC coded system with soft-output MMSE detection.}
\label{MF_BPSK}
\end{figure}

Figure \ref{MF_BPSK} shows the performance comparison between the NBLDPC coded systems with the proposed soft output MF-based detector
and the NBLDPC coded systems with soft output MMSE detector described in the previous section when the number of transmit antennas is $N_t=200$.
The rates of NBLDPC codes are ranging from 1/12 to 1/3 and the information length is 800 bits ($K=100$ symbols).
As mentioned in Section III, we select the low rate NBLDPC codes 
since we expect that the operating region, e.g. BER of $10^{-4}$, will be occurred in low SNRs 
at which the proposed detection is near optimal.
For $R=1/3$, the performance of coded system with MMSE detector is better than 
that of the proposed system by about 0.8 dB.
This is because the uncoded performance of MF-based detector is worse than that of MMSE detector by roughly 2 dB 
(not show here).
However, as we expected, the performance difference between two coded systems (one with MMSE detector and another one with the proposed detector) is vanished for $R=1/6,1/9$ and $1/12$
since, in low SNRs, both detectors provides almost the same detecting performance.
Thus, the soft output generation from MF detector presented in this paper works well with NBLDPC decoder.
Note that the codes with $R < 1/3$ are constructed from NBLDPC of rate $1/3$ 
by using the instruction given in \cite{MRNB}.

\begin{figure}[htb]
\centering
\includegraphics[scale=0.7]{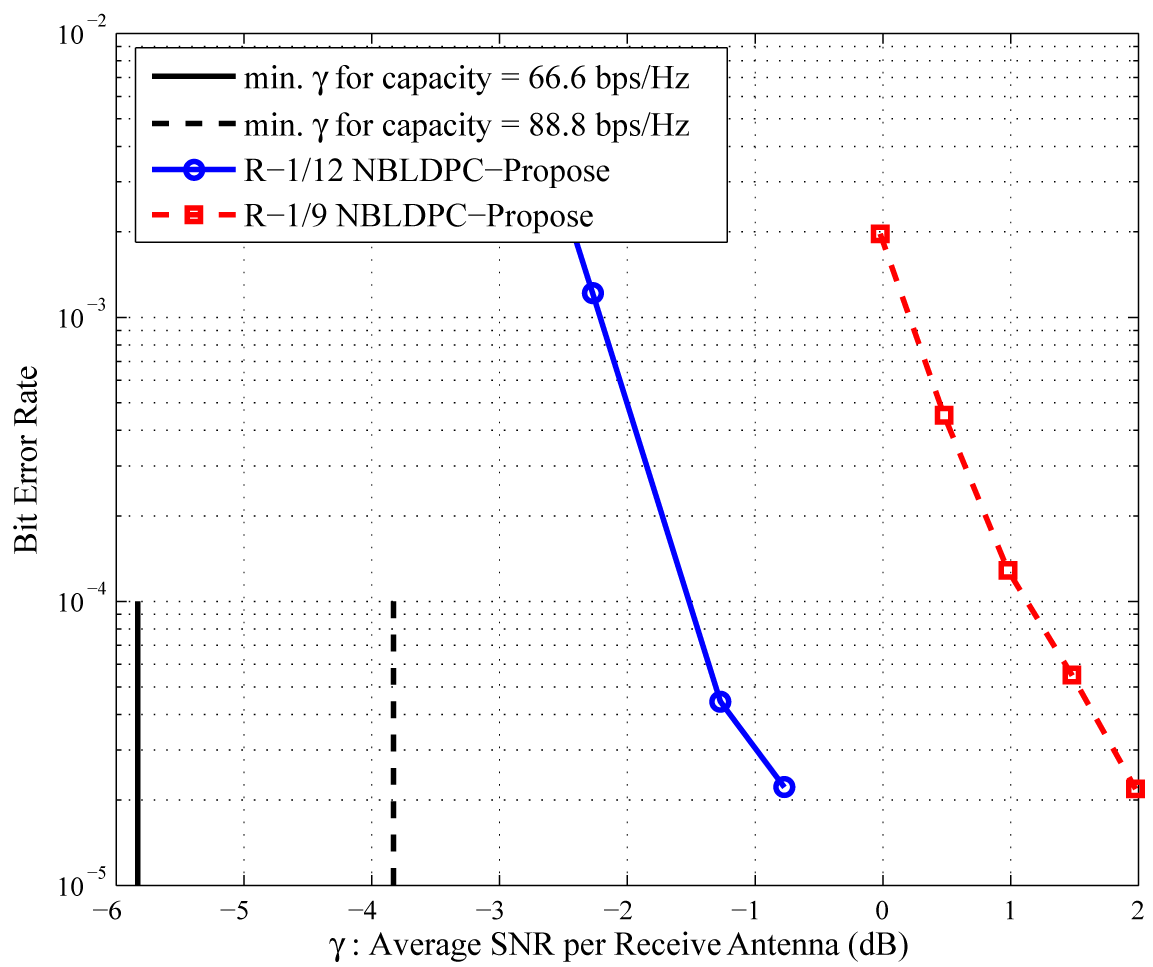}
\caption{BER of the NBLDPC coded $200 \times 200$ MIMO systems with the proposed soft output MF-based detector and 16-QAM modulation.
The information size $\mathrm{k}=800$ bits and the code length is $800/R$ bits.}
\label{MF_16QAM}
\end{figure}

We have shown in Fig. \ref{MF_BPSK} that
the application of the proposed soft output MF-based detector in NBLDPC coded MIMO system with binary modulation is excellent if the rate of NBLDPC code is low, e.g., $R < 1/3$.
The next step is to illustrate the coded performance with higher order modulation.
Figure \ref{MF_16QAM} shows the performance of NBLDPC codes of $R=1/9,1/12$ when the 16-QAM modulation is employed.
Unlike the case of BPSK modulation,
we can see the error floor at BER of $10^{-4}$.
The error floor appears because the proposed detection has no interference cancellation which is more severe to
the case of higher order modulation (more points in modulation constellation comparing with the binary case).
However, we observed that the number of bit errors in each frame error is small.
For example, 100 frame errors at $\gamma = 2$ dB for NBLDPC code of $R=1/9$ equal to 1752 bits error.
The average bit errors per one frame error is just 18 bits which is rather small.
We intend to use the joint detection-decoding to overcome the error floor
but this topic is left as the future work.

In this part, we have demonstrated the application of the proposed soft output MF-based detector in NBLPDC coded large MIMO systems.
The proposed detector can be efficiently used with both binary modulation but we mention that this kind of detector is useful
in the limited operation region, i.e., the region that MF detector is near optimal.
In that region, we need to use the low rate channel code and the low rate NBLDPC codes are the promising candidate as we shown in Fig. \ref{MF_BPSK}.
Please note that the NBLDPC codes of $R<1/3$ used in this work 
can be decoded at almost the same computational complexity \cite{MRNB}
by using the Tanner graph of NBLDPC code of $R=1/3$
which is another advantage of using NBLDPC codes.

\subsection{Complexity Analysis}
Following \cite[p.~18]{complexity_cal,matrix_computation}, 
the complexity of complex matrix calculation is expressed in terms of complex floating point operations (flops).
The number of flops for some mathematical operations is given in Table \ref{matrix_comp}.
We note that the detection for coded MIMO systems comprises of two steps : 1) detection and 2) soft output generation.
\begin{table}[htb]
\setbox0\hbox{\verb/\documentclass/}%
\caption{The number of flops for some operations \cite{complexity_cal} and \cite[p.~193]{numerical_comp} for exponential calculation where 
 $a,b \in \mathbb{R}$, $\mathbf{a},\mathbf{b} \in \mathbb{C}$, and $\mathbf{c},\mathbf{d} \in \mathbb{C}^{N_r \times 1}$}
\label{matrix_comp}
\begin{center}
\begin{tabular}{| c | c |  c | }
 \hline
 Operation (name) & Operation (math.) &  Number of flops \\
  \hline
Real multiplication & $ab$ & 1 \\
Complex multiplication & $ \mathbf{a}\mathbf{b}$ & 3 \\
Real addition & $a+b$ & 1 \\
Complex addition & $\mathbf{a} +\mathbf{b}$ & 1 \\
Inner product & $\mathbf{c^{\mathsf{H}}}\mathbf{d}$ & $4N_r-1$ \\
Scalar-vector multiplication & $\mathbf{a} \cdot a$ & $N_r$ \\
Exponential calculation & $\text{exp}(a)$ & 50 \\
  \hline
\end{tabular}
\end{center}
\end{table}

\subsubsection{Proposed Soft Output MF-based Detector \cite{MF_large_MIMO}}
The estimation of $\hat{s}_k$ as defined in (\ref{detect_eq}) needs two steps : 
1) Computing the weight matrix  $\mathbf{W}_k$ according to (\ref{reduc1}) which requires $N_r$ flops.
2) Multiplying $\mathbf{W}_k$ with received vector $\mathbf{y}$ which requires $4N_r-1$ flops.
Since we have $N_r$ receive antennas, 
the total flops used for detection are $5N_r^2 - N_r$.
Thus, the per-bit complexity is $\mathcal{O}(N_r)$.
To obtain the soft output defined in (\ref{soft_MF}),
we need one flop for computing squared euclidean norm,  one  flop for subtraction, $3$ flops for multiplication of real constant, 
Each transmitted symbol has $M$ levels to be calculated and we have $N_t = N_r$ transmitted symbols.
Therefore, in total, $55MN_r$ flops are required for generating soft outputs. 
In summary, the overall per-bit complexity for both detection and generating soft output is $\mathcal{O}(N_r)$.

\subsubsection{Soft Output MMSE Detector}
In the case of MMSE detector, it has been shown that the complexity for detection 
is $10N_r^3 + 5.5N_r^2 + 1.5N_r$ flops \cite{complexity_cal}. 
To generate the soft outputs as defined in (\ref{Soft_MMSE}),
$\mu_k = \mathbf{W}_{\textrm{mmse},k}^{\mathsf{H}}\mathbf{H}_k$ requires $4N_r - 1$ flops and  
$\epsilon^{2}_{k} = \frac{E_s}{N_t} ( \mu_k - \mu_{k}^{2} )$ requires 3 flops.
Again, each transmitted symbol have $M$ level to be calculated and we have $N_t = N_r$ transmitted symbols.
So, $4MN_r^2 + 58MN_r$ flops 
are required for generating soft outputs from all estimated symbols.

\begin{figure}[htb]
\centering
\includegraphics[scale=0.7]{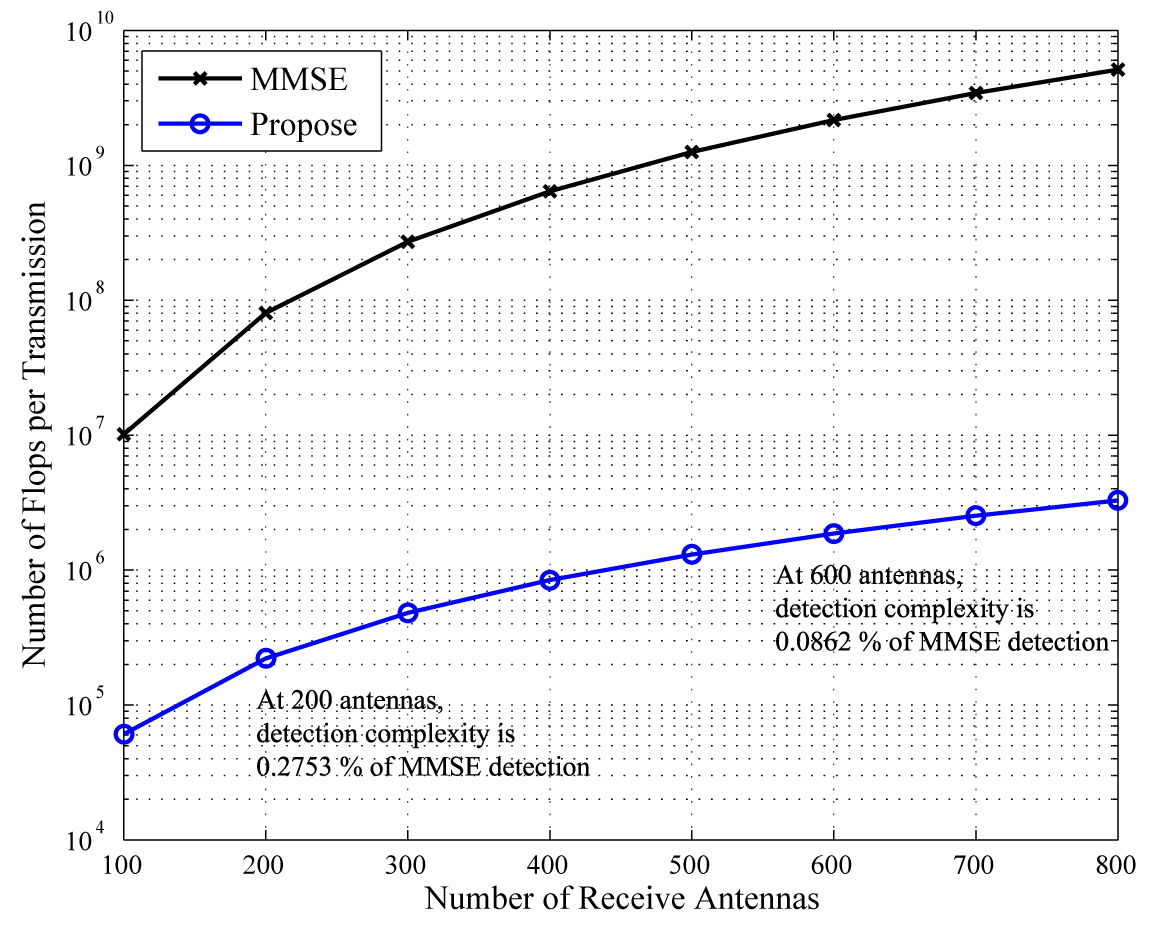}
\caption{The computational complexity of the MMSE detection and the proposed detection versus the number of receive antennas.}
\label{Complexity_Comparison}
\end{figure}

Table \ref{flop_need} summarises the computational complexity of both the proposed detector and the MMSE detector in terms of flops.
In the case of BPSK modulation ($M=2$), 
the computational complexity versus the number of receive antennas is plotted in Fig. \ref{Complexity_Comparison}.
It is seen from the figure that the computational complexity of the proposed detection is much lower than that of MMSE detection
For instance,
$80,563,500$ flops are required for the MMSE detector while
we need only $221,800$ flops for the proposed MF detector  (just 0.28 $\%$ to the MMSE detection)
Thus, by using the proposed detector, the computational complexity of large MIMO detection is significantly reduced.
\begin{table}[htb]
\setbox0\hbox{\verb/\documentclass/}%
\caption{The number of flops for MIMO detection.}
\label{flop_need}
\begin{center}
\begin{tabular}{| c | c |  c | }
 \hline
 Type $\diagdown$ Operation & Detection &  Soft Output \\
  \hline
Proposed detector & $5N_r^2 - N_r$ & $55MN_r$ \\
  \hline
MMSE detector & $10N_r^3 + 5.5N_r^2 + 1.5N_r$  & $4MN_r^2 + 58MN_r$ \\
  \hline
\end{tabular}
\end{center}
\end{table}

\section{Threshold of Non-Binary LDPC Codes in Large MIMO Systems}
In the previous two sections,
the performances of NBLDPC coded large MIMO systems
with both the MMSE detection and the proposed MF-based detection have been investigated
by using the practical code lengths, e.g., a few thousand bits.
We found that NBLDPC coded systems have the capacity-gaps about 3.5 dB and 6 dB, which can be considered as
the best performing for BPSK modulation and 16-QAM, respectively.
The fact from coding theory tells us that the capacity-gap can be reduced by using the larger code length as demonstrated in Fig. \ref{ber600x600_16QAM}.
So, one may think about the performance of NBLDPC codes of large code lengths.
We now show in this section the capacity-gap of NBLDPC coded large MIMO systems which have very large code length.

In this section, we analyse the performance of NBLDPC coded large MIMO systems in the limit of infinite code length
by using the Monte Carlo density evolution proposed in \cite{Davey_Thesis}.
The Monte Carlo density evolution is a tool for calculating the minimum SNR 
(known as decoding threshold) at which the NBLDPC codes can reliably decode the noisy received symbols.
The calculating steps of Monte Carlo density evolution can be summarized here. 

\noindent1) Input the size of variable node $L$, number of iteration $\ell_{max}$, 
and the initial SNR $\gamma^{(0)}$ in dB.
The size of variable node $L$ must be large enough to represent the infinite code length, e.g., $10^5$ nodes.
The maximum iteration $\ell_{max}$ is also large, e.g., 2000 iterations.
The initial SNR $\gamma^{(0)}$ should start from the value that NBLDPC decoder can reliably decode.,
e.g., the SNR that can provide BER of $10^{-4}$ in the case of practical code length.
Let $\ell$ be the index of iteration.

\begin{figure}[htb]
\centering
\includegraphics[scale=0.7]{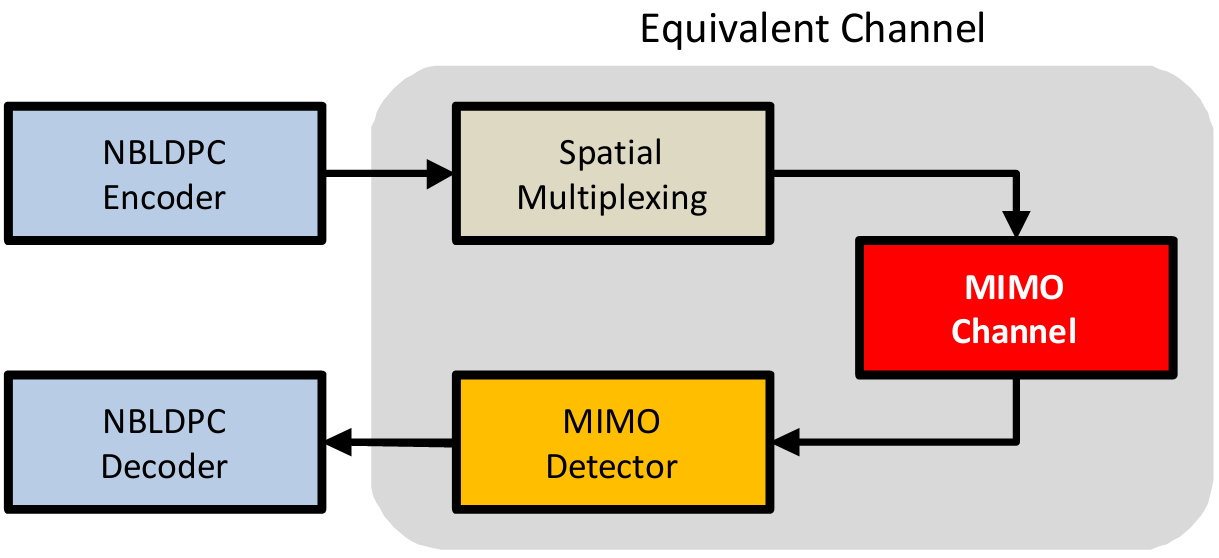}
\caption{An equivalent channel of NBLDPC coded MIMO systems.}
\label{Eq_Channel}
\end{figure}

\noindent2) By assuming the zero codeword to represent the NBLDPC codeword,
we first produce $L$ noisy variable nodes with associated likelihoods $p^{(0)}_v(x_j)$
from the  equivalent channel as shown in Fig. \ref{Eq_Channel}
where $v=1,\ldots,L$ and $x_j \in \GF(2^m), j=1,\ldots,2^m$.

\begin{figure}[htb]
\centering
\includegraphics[scale=0.8]{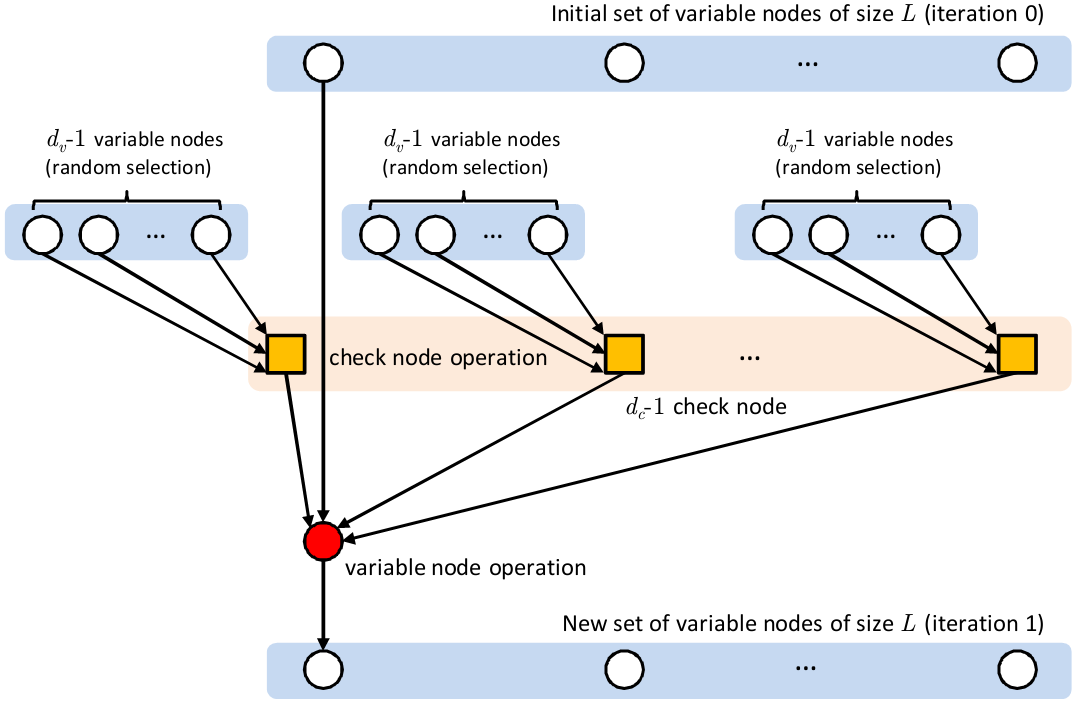}
\caption{The generation of new set (ensemble) of variable nodes in the Monte Carlo analysis 
originally proposed by M. Davey \cite{Davey_Thesis}.
By iterating the computation to certain iteration (quite large),  this is equivalent to perform decoding on cycle free Tanner graph with infinite size of variable nodes.}
\label{MCDE}
\end{figure}

\noindent3) Increment $\ell = \ell+1$.
Creating the new set of variable nodes of  size $L$ with the associated likelihood $p^{(\ell)}_v(x)$
by using the concept of BP decoder \cite{nb_ldpc}.
The operation in this step, which is illustrated in Fig.\ref{MCDE}, is like running BP algorithm on cycle-free Tanner graph.
The full detail of this step  can be found in \cite[p.22]{Davey_Thesis}.

\noindent4) Replacing the previous set of variable nodes of size $L$ with the new set of variable node of the same size.
Calculating the average Shannon entropy as shown below
\begin{equation}
\label{Prob_eq}
H =  - \frac{1}{L}\sum\limits_{v = 1}^L {\sum\limits_{j = 1}^{2^m } {p_v^{(l)} \left( {x_j } \right)} } \log _{2^m } \left( {p_v^{(\ell)} \left( {x_j } \right)} \right),
\end{equation}
where, in this case, the value of $H$ can be used to represent the ambiguity of the underlying variable nodes.
Thus, the value of $H=0$ implies a complete removal of the ambiguity.
In this paper, we choose $H_{\rm{stop}}=10^{-6}$ as the stopping criterion.
If $H>H_{\rm{stop}}$ and $\ell < \ell_{max}$, repeat step 3 and step 4 in iterative manner until $\ell = \ell_{max}$.
If $H \leq H_{\rm{stop}}$, go to step 5.
If $H>H_{\rm{stop}}$ and $\ell = \ell_{max}$, go to step 6.

\noindent5) If $H \leq H_{\rm{stop}}$, this means the NBLDPC decoder can reliably decode the noisy variable nodes at $\gamma^{(\ell)}$.
Decrease the $\gamma^{(\ell)}$ by small step size, e.g., $\gamma^{(\ell+1)} = \gamma^{(\ell)} -0.05$ dB
and then repeat step 2 to step 4.

\noindent6) If $H>H_{\rm{stop}}$ and $\ell = \ell_{max}$, this means the NBLDPC decoder cannot reliably decode the noisy variable nodes at $\gamma^{(\ell)}$.
Declare the previous $\gamma^{(\ell-1)}$ as the decoding threshold.

\begin{figure}[htb]
\centering
\includegraphics[scale=0.75]{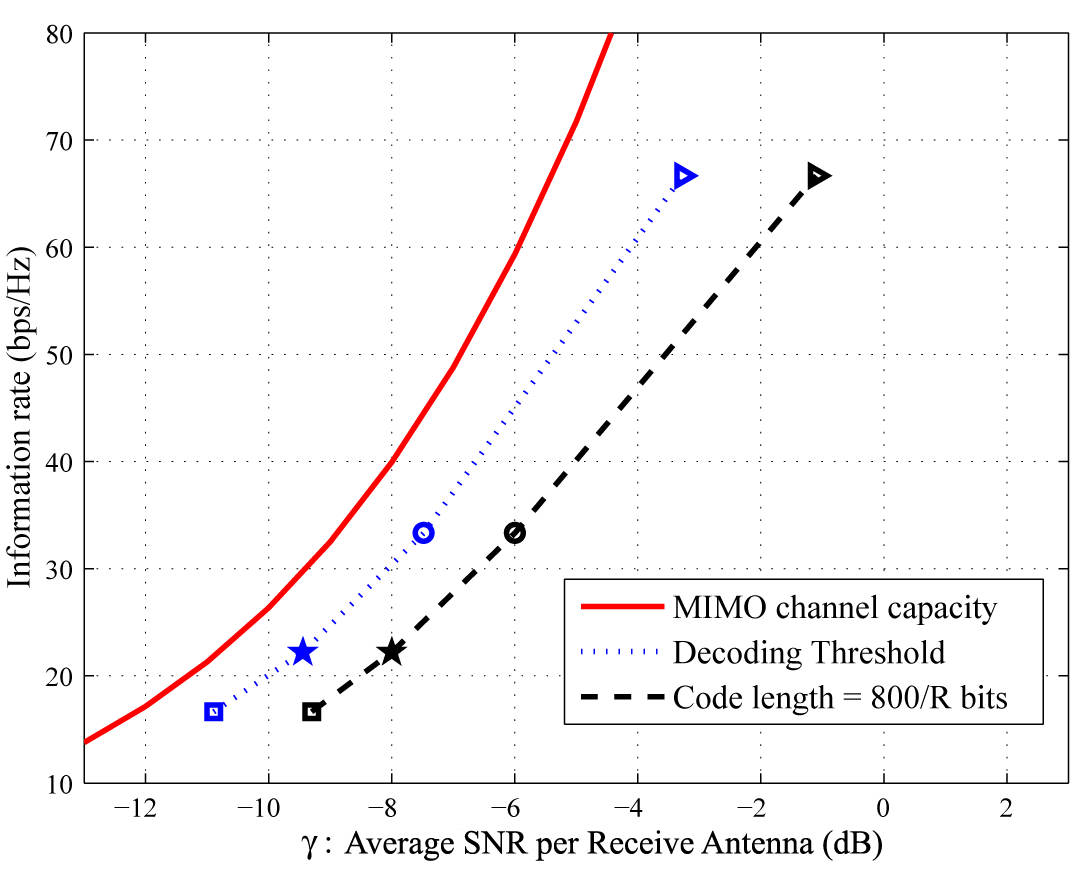}
\caption{The decoding threshold of NBLDPC coded $200 \times 200$ MIMO system with BPSK modulation.
The corresponding MIMO capacity and the performance at code length $n=800/R$ bits are also plotted.}
\label{MCDE_result_200x200}
\end{figure}

Figure \ref{MCDE_result_200x200} shows
the decoding threshold of NBLDPC coded $200 \times 200$ MIMO system with the proposed MF-based detection and BPSK modulation.
We also plotted the MIMO capacity of  $200 \times 200$ MIMO system and 
the SNR points required to reach the BER of $10^{-4}$
of the NBLDPC coded system with short information length $k=800/R$ bits.
As you can see from the figure, the capacity-gap of NBLDPC codes does not go to small value even when the code length tends to infinity.
The capacity-gap of NBLDPC codes to the MIMO capacity is about 1.6 dB for all rates.
Therefore, we cannot achieve the MIMO capacity by simple concatenation the large (spatial multiplexing) MIMO systems 
with regular NBLDPC codes over $\GF(2^8)$. 
The methodologies to reduce the remaining capacity-gap are challenging and will be explored in the future work.

\begin{figure}[htb]
\centering
\includegraphics[scale=0.75]{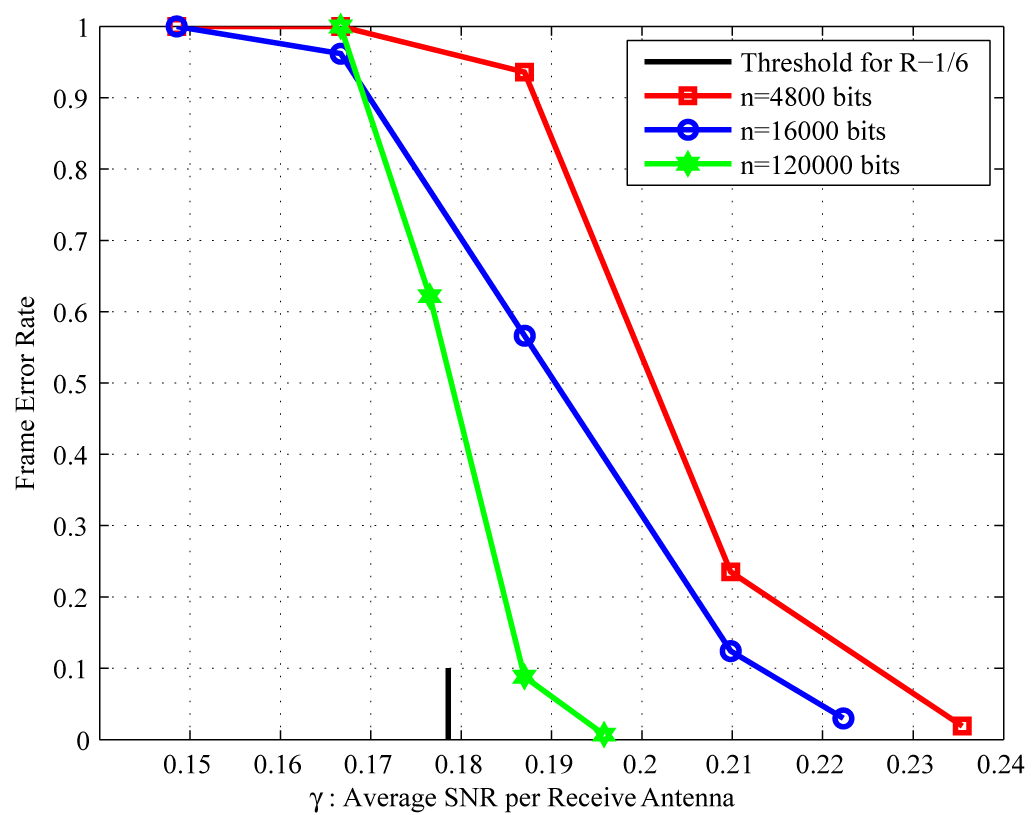}
\caption{Frame error rate of NBLDPC codes of $R=1/6$ over the $200 \time 200$ MIMO channel with BPSK modulation.}
\label{MCDE_confirm}
\end{figure}

We also confirm the correctness of the decoding threshold obtained from Monte Carlo analysis
by conducting the simulation of the code with very long code length.
It can be observed from Fig. \ref{MCDE_confirm} that 
the frame error rate converges to the decoding threshold as the code length increases.

We note that the same Monte Carlo density evolution cannot be used to 
compute the decoding threshold of NBLDPC coded MIMO systems with $M$-QAM modulation where $M>2$.
This is because the $M$-QAM modulation does not has the property of  rotational symmetry
so the zero codeword cannot be used to represent the LDPC codeword.
The modified Monte Carlo density evolution presented in \cite{Rong_GFq_Qmod} can be used to 
compute the decoding threshold of NBLDPC coded MIMO systems with $M$-QAM.
However, this method is computational intensive so we leave 
the decoding threshold of NBLDPC coded MIMO systems with $M$-QAM as the future work.

\section{Effect of Channel Estimation Errors}
Beside high detection complexity, 
channel estimation is also a major bottleneck in large MIMO systems \cite{marzetta_blast_training}.
In this section, the effect of imperfect channel estimation on the NBLDPC coded large MIMO systems will be investigated.
After performance evaluating, we expect to answer the readers about 
the sensitivity of NBLDPC coded large MIMO systems
to the accuracy of channel estimation.
By using the model adopted in \cite{coded_large_mimo1}, 
the estimated channel matrix to be used at the receiver is given by
\begin{equation}
\label{CE_model}
{\bf \tilde H} = {\bf H} + \Delta {\bf H},
\end{equation}
where $\Delta {\bf H}$ is the estimation error matrix in which the entries are assume to be the complex Gaussian with zero mean and variance $\sigma^{2}_{e}$.

\begin{figure}[htb]
\centering
\includegraphics[scale=0.65]{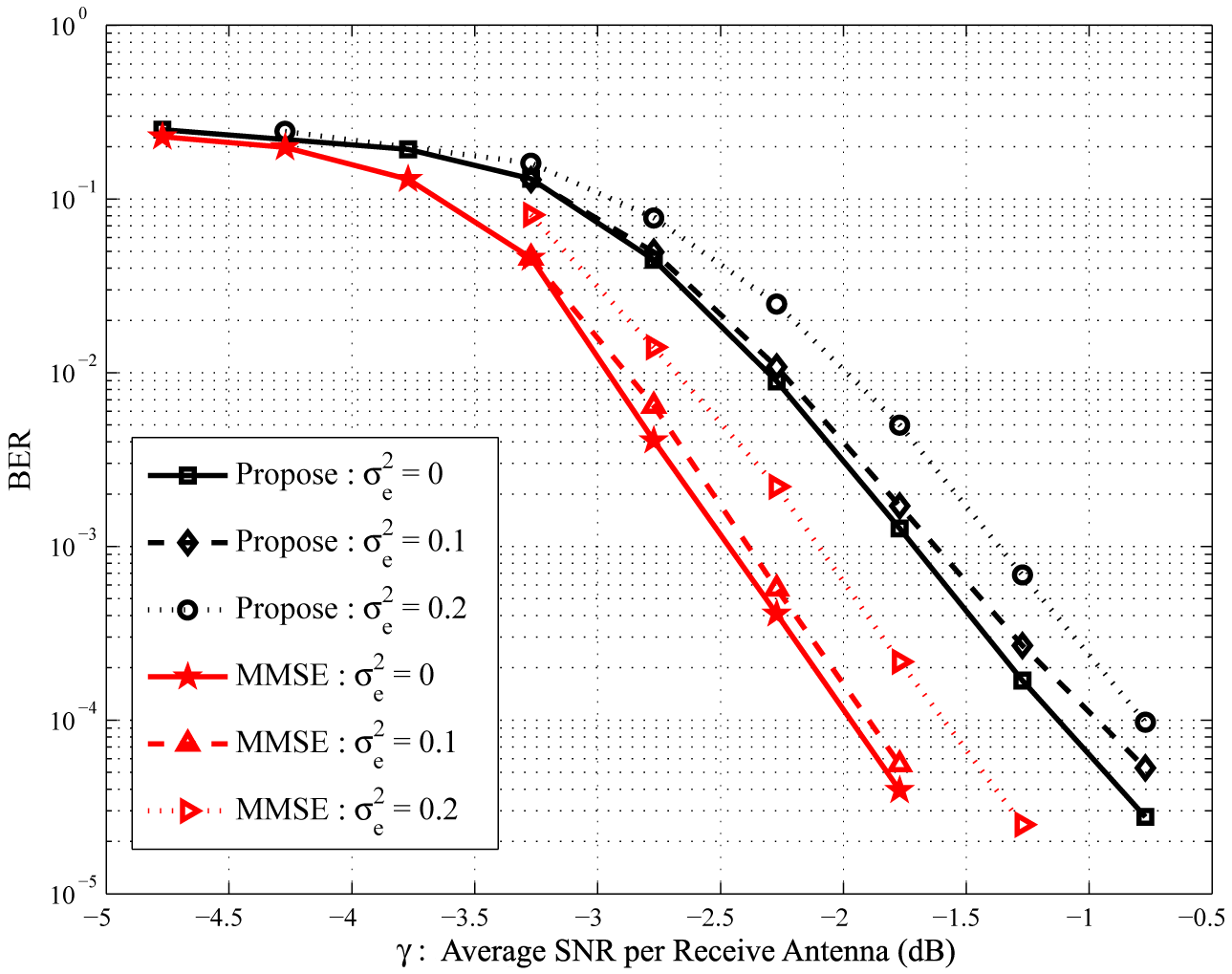}
\caption{BER of the NBLDPC coded $200 \times 200$ MIMO systems with BPSK modulation for different values of $\sigma^{2}_{e}$ by which $\sigma^{2}_{e}=0$ means perfect channel estimation.
In this figure, the NBLDPC code of $R=1/3$ and $\mathrm{n}=2400$ bits is employed
``Propose" means the non-binary LDPC coded system with the proposed soft output MF-based detection and ``MMSE" represents the non-binary LDPC coded system with soft-output MMSE detection.}
\label{Eff_CE_R13}
\end{figure}

\begin{figure}[htb]
\centering
\includegraphics[scale=0.82]{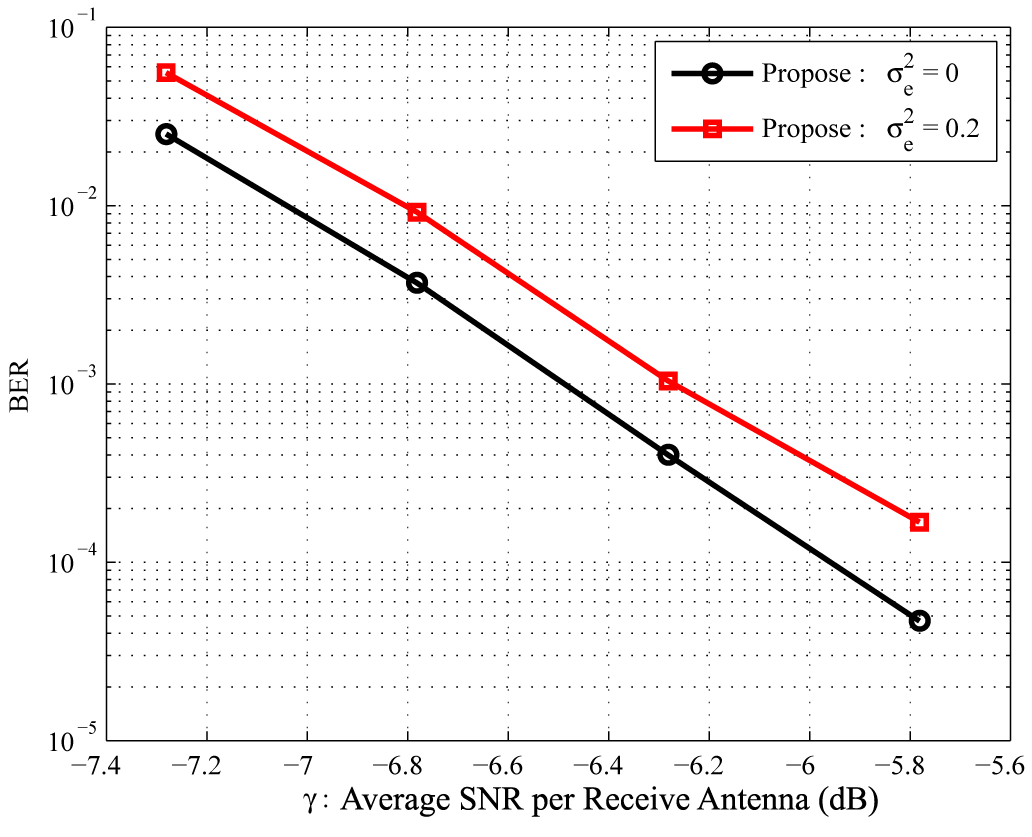}
\caption{BER of the NBLDPC coded $600 \times 600$ MIMO systems with BPSK modulation for different values of $\sigma^{2}_{e}$.
In this figure, the NBLDPC code of $R=1/6$ and $\mathrm{n}=4800$ bits is employed
``Propose" means the NBLDPC coded system with the proposed soft output MF-based detection.}
\label{Eff_CE_R16}
\end{figure}

Figure \ref{Eff_CE_R13} shows
the BER of the NBLDPC coded $200 \times 200$ MIMO systems with BPSK modulation for different values of $\sigma^{2}_{e} = 0,0.1,0.2$.
For both the MMSE and the proposed detections, the coded performance degrades when $\sigma^{2}_{e}$ increases as expected, 
e.g., worse channel estimation quality, worse coded performance.
However, the figure clearly shows that 
the performance degradation is negligible even at large error variance $\sigma^{2}_{e}=0.1$
(the error variance at this level is considered to be large  \cite{Park_channel_est}).
Note that the performance with $\sigma^{2}_{e} < 0.1$ is identical to that of $\sigma^{2}_{e} = 0.0$ 
so we do not included in the figure.
For larger error variance $\sigma^{2}_{e}=0.2$,
the performance degradation is only 0.4 dB which is still small.
The same property can also be seen although the channel dimension is increased from 200 to 600 as shown in Fig. \ref{Eff_CE_R16}.

The channel estimation algorithm that can render the error variance to be $0.2$
is enough to provide not too large performance loss.
The reasons why the effect of channel estimation errors to NBLPDC coded large MIMO systems is quite small are :
1) the effect of channel estimation errors for uncoded performance in near capacity region is small,
2) the powerfulness of NBLDPC codes,
3) if one look at the simplification of the proposed MF-based detection,
one may think that this detection scheme does not require the perfectness of channel information at receiver side. 
Based on the results presented in this section,
it is sufficient to conclude that the NBLDPC coded large MIMO systems with MMSE and proposed detections
are very robust to channel estimation errors.

\section{Effect of Spatial Correlation}

All the previous results of NBLDPC coded large MIMO systems have been evaluated by assuming i.i.d. flat fading channel, 
i.e., each entry of $\bf{H}$ is modelled as identically and independent complex Gaussian random variable.  
However, the realistic MIMO channels have spatial correlation, i.e., the entries of $\bf{H}$ (the propagation coefficients) are correlated, 
which has the adverse effect on the MIMO capacity as well as the error rate.
It is known that low angle spread, insufficient antenna spacing and lack of rich scattering may cause the spatial correlation between antennas \cite{hedayat_analysis_STC_correlate}.
For large MIMO systems, the insufficient antenna spacing is an important factor to cause the spatial correlated MIMO channel
since a large number of antennas, e.g., 200 transmit/receive antennas, need to be placed in a limited space. 
So, the study of spatial correlation in NBLDPC coded large MIMO systems should be addressed.

Since, in this work, both the transmitter and the receiver have large number of antennas,
we need to consider the doubly correlated MIMO channels models which can be written as \cite{shin_C_doubly}
\begin{equation}
\label{DC_MIMO_model}
{\bf H} = {\bf R }_r^{1/2} {\bf H}_{i.i.d.} {\bf R }_t^{1/2}, 
\end{equation}
where ${\bf H}_{i.i.d.}$ is the channel fading matrix of size $N_r \times N_t$
whose entry is assumed to be i.i.d. complex Gaussian random variable with zero mean and unit variance, 
${\bf R }_r$ is the received correlation matrix of size $N_r \times N_r$, 
and ${\bf R}_t$ is the transmitted correlation matrix of size $N_t \times N_t$.

In fact, there are many ways to generate the correlation matrices ${\bf R }_r$ and ${\bf R}_t$.
To study the effect of correlation on the MIMO systems, 
we first consider an simple and well-known exponential correlation model as shown below
\begin{equation}
\label{C_coeff}
\begin{array}{*{20}c}
{{\bf R}_r \left( {\rho_r } \right) = \left[ {\rho_r^{\left| {i - j} \right|} } \right]_{i,j = 1, \ldots ,N_r } , ~ \rho_r  \in \left[ {0,1} \right)},\\
{{\bf R}_t \left( {\rho_t } \right) = \left[ {\rho_t^{\left| {i - j} \right|} } \right]_{i,j = 1, \ldots ,N_t } , ~ \rho_t  \in \left[ {0,1} \right)},\\
\end{array}
\end{equation}
where $\rho_t$ and $\rho_t$  denotes the transmitted and received correlation parameter, respectively.
This model is reasonable for equally-spaced linear antenna array \cite{shin_C_doubly}.
Furthermore, this model is physically reasonable in the sense that the correlation decreases 
with the increasing of distance between transmit/receive antennas \cite{Loyka_expo_mimo}.

\begin{figure}[htb]
\centering
\includegraphics[scale=0.7]{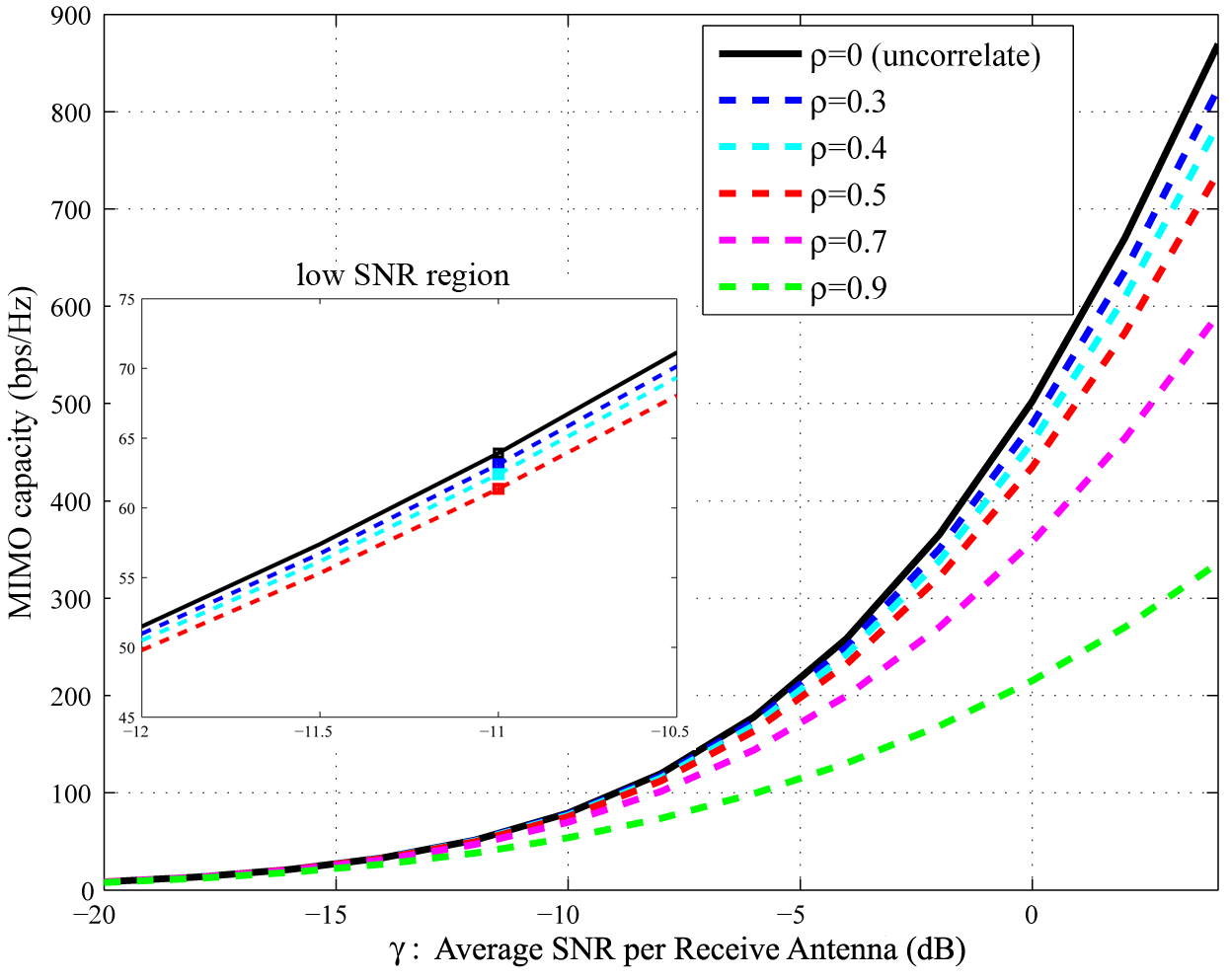}
\caption{Capacity of $600 \times 600$ MIMO system by using an exponential correlation model with different correlation parameters $\rho = \rho_r=\rho_t$.}
\label{correlated_MIMO_capacity}
\end{figure}

Figure \ref{correlated_MIMO_capacity} shows the effect of correlation on MIMO capacity when $N_t=N_r=600$ and the modulation is BPSK.
We denote parameter $\rho = \rho_t = \rho_r$ as the correlation parameter which is the same for both the transmitter and the receiver.
The parameter $\rho=0$ means uncorrelated system whereas $\rho = 1$ means fully correlated system.
It is not surprising to see that the capacity of MIMO systems decreases with the increasing of correlation parameters.
This is because the spatial correlation 
has the adverse effect on rank structure of $\mathbf{H}$ which directly relates to the MIMO capacity.
Increasing the number of receive antenna is found to be 
the solution to keep the MIMO capacity as well as the coded performance at the same level \cite{coded_large_mimo3}.

\begin{figure}[htb]
\centering
\includegraphics[scale=0.7]{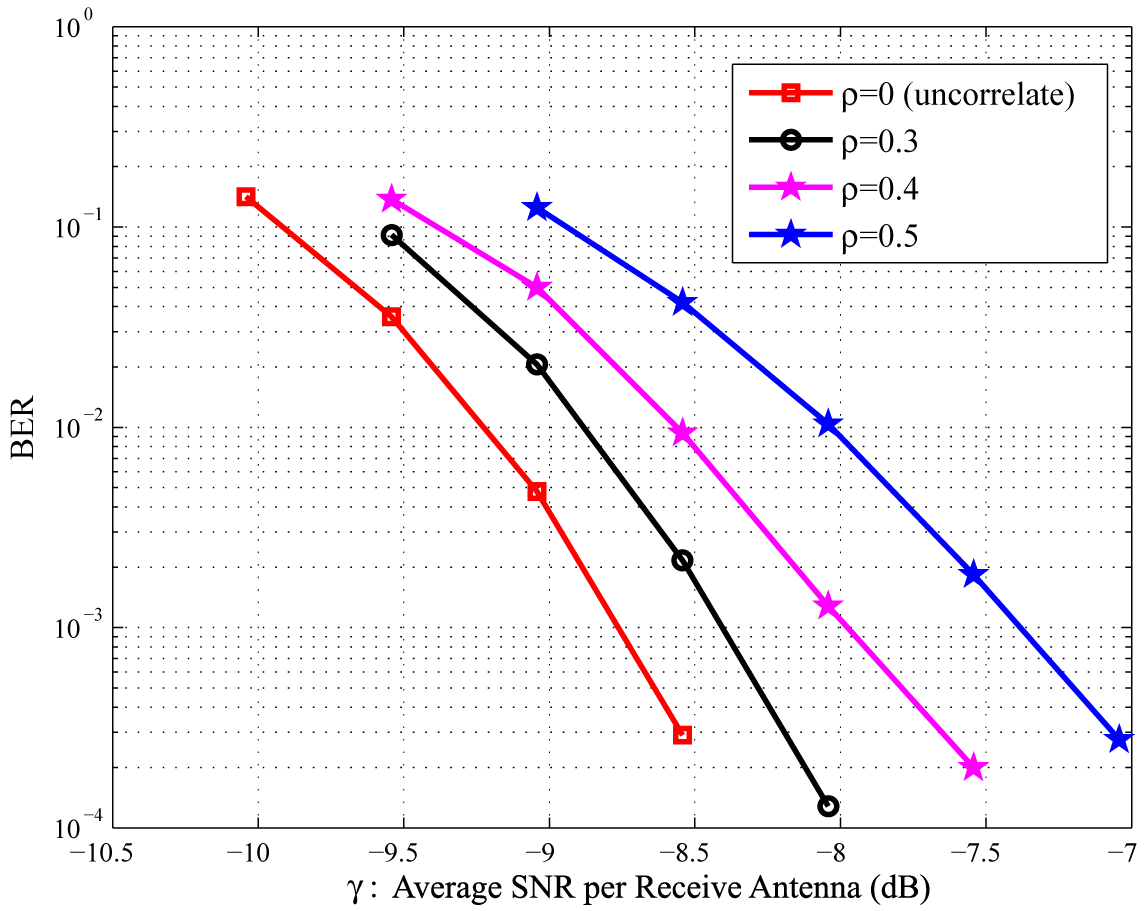}
\caption{Performance of $R=1/9$ NBLDPC coded $600 \times 600$ MIMO systems in the presence of correlation (exponential correlation model).
BPSK and the proposed detection are utilized as modulation scheme and detector, respectively.
The spectral efficiency is about 66.67 bps/Hz and the information length is $\mathrm{k}=800$ bits.}
\label{correlated_MIMO_BER}
\end{figure}

Fortunately, 
one can expect small performance degradation in low SNR region 
since the effect of correlation in low SNR region is quite small.
For example, the capacity at $\gamma = -11$ dB for uncorrelated case is about 64 bps/Hz.
We observe that the loss in capacities at $\gamma = -11$ dB for $\rho = 0.3,0.4,0.5$ 
are just 0.85, 1.5, 2.7 bits, respectively, which are rather small if one look at the loss in high SNR region.
At low SNR region, we know that the proposed MF-based detection which has very low computational complexity
can be efficiently utilized.
To illustrate this, we show the performance of NBLDPC coded $600 \times 600$ MIMO systems at spectral efficiency 66.67 bps/Hz in Fig. \ref{correlated_MIMO_BER}.
Comparing to uncorrelated case $\rho = 0$,
the performance loss from $\rho= 0.3,0.4,0.5$ are approximately 0.4,1,1.5 dB, respectively.
If the bandwidth is 15 MHz (not too large bandwidth requirement),
the NBLDPC coded large MIMO systems that suffer 
from the spatial correlation could be the promising candidate to provide
the the reliable communication (BER of $10^{-4}$) with data rate 1 Gbps (very high) 
and the received power around -8 dB (very low).

\begin{figure}[htb]
\centering
\includegraphics[scale=0.7]{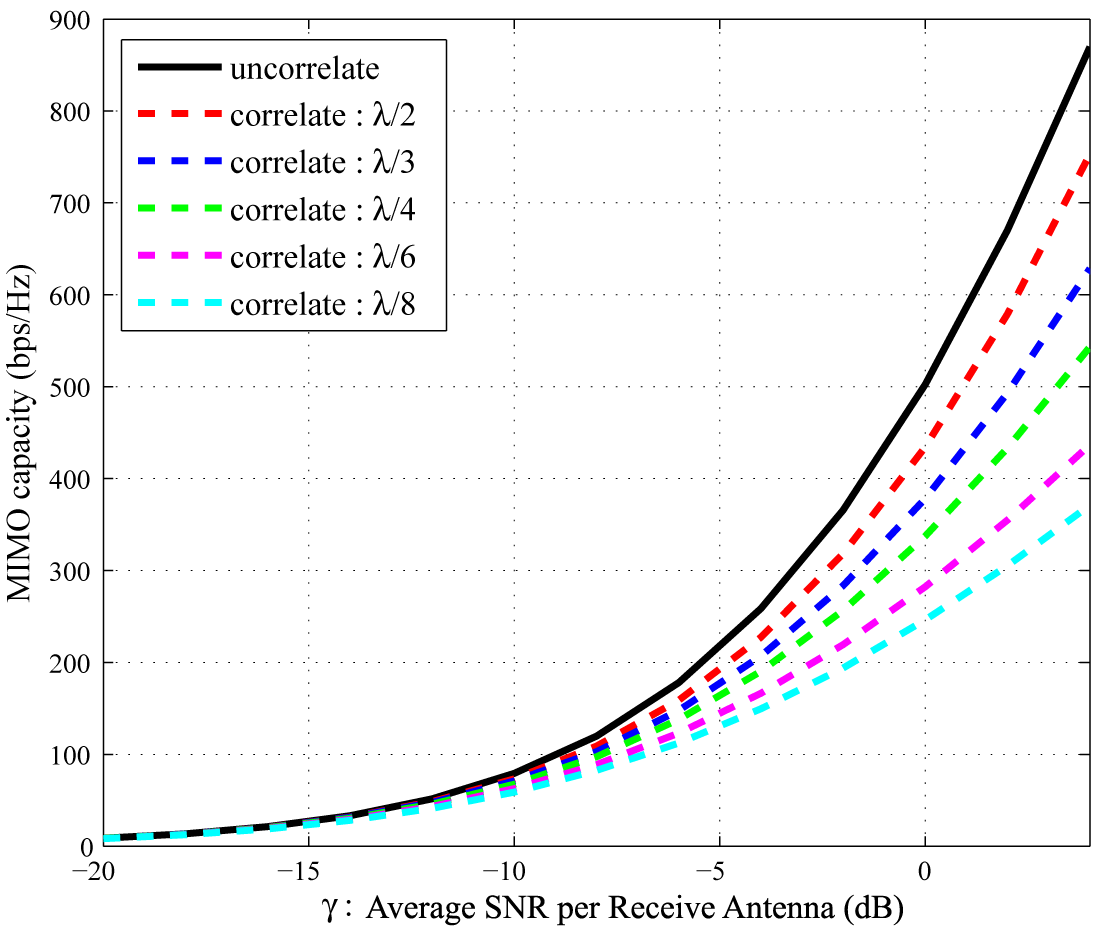}
\caption{Capacity of $600 \times 600$ MIMO system under correlated environment presented in \cite{thushara_corr} 
with different values of antenna spacing.
By using antenna array configuration, a number behind ``correlated :" indicates the antenna spacing between two adjacent antennas.
The carrier frequency is set to 5 GHz which corresponds to  wavelength $\lambda = 6$cm.
$\phi_{0} = \pi/2$ and $\Delta \phi=\pi/6$(see the definition of these parameters in \cite{thushara_corr}.) 
are used to generate the correlation coefficients}
\label{correlate_MIMO_thushara}
\end{figure}

Fig. \ref{correlate_MIMO_thushara} shows the MIMO capacity under correlated environment modelled by the instruction given in \cite{thushara_corr}.
This model allows us to study the effect of antenna spacing which is of interest in large MIMO systems.
The MIMO capacity decreases if the antenna spacing decreases as seen from Fig. \ref{correlate_MIMO_thushara}.
Like the exponential correlation model that we have previously discussed, 
there still exists the points in low SNR region, at which the effect of spatial correlation is rather small. 
So, at that region, we can utilize the NBLPDC coded system with the proposed ultra low-complexity MF-based detection to construct
the excellent MIMO communication although the antenna spacing is less than half of wavelength.

\section{Conclusions}
In this work, we have demonstrated the excellent performance of $(2,d_c)$- regular NBLDPC codes over $\GF(2^8)$
in the MIMO systems with hundreds of antennas.
The advantage of using NBLPDC codes in these large MIMO systems can be listed here : 
1) very good error rate by using the low-complexity detections,
2) robust to channel estimation errors,
3) application in correlated environment is excellent.
We also proposed in this work the ultra low-complexity soft output MF-based detection which can be efficiently utilized
for NBLDPC coded systems in near capacity region.
Supremely, we believe that the results presented in this work can be used as the benchmark for the further study on coded large MIMO systems.

\section*{Acknowledgments}
This work is financially supported by the Telecommunications Research Industrial and Development Institute (TRIDI), with National Telecommunications Commission (NTC), Grant No.PHD/009/2552. 
The authors would like to thank anonymous reviewers of ISIT2012 and ISTC2012 for their helpful comments and suggestions.

\bibliographystyle{IEEEtran}
\bibliography{IEEEabrv,my_references}

\end{document}